\documentclass[11pt]{article}
\usepackage[margin=1in]{geometry}
\usepackage{amsmath,amssymb}
\usepackage{booktabs}
\usepackage{tabularx}
\usepackage{array}
\usepackage[table]{xcolor}
\usepackage{tikz}
\usetikzlibrary{positioning,arrows.meta,calc,decorations.pathreplacing,shapes.geometric,fit,backgrounds,quantikz2}
\usepackage{pgfplots}
\pgfplotsset{compat=1.18}
\usepgfplotslibrary{fillbetween}
\usepackage[section]{placeins}
\usepackage{caption}
\usepackage[numbers,sort&compress]{natbib}
\usepackage{xurl}
\usepackage[colorlinks=true,linkcolor=cC,citecolor=cC,urlcolor=cC]{hyperref}

\definecolor{cR}{HTML}{E4002B}   % arm R, random assignment (red)
\definecolor{cC}{HTML}{0072CE}   % arm C, natural pipeline (blue)
\definecolor{cN}{HTML}{00A651}   % arm N, null circuit (green)
\definecolor{cK}{HTML}{FF8C00}   % classical features (orange)
\definecolor{cQ}{HTML}{8E44AD}   % noiseless features (purple)
\definecolor{cT}{HTML}{00B5B8}   % teal
\definecolor{cM}{HTML}{E6007E}   % magenta
\definecolor{cY}{HTML}{F5B800}   % gold
\definecolor{cG}{HTML}{6B7280}   % gray

\newcommand{\codeDOI}{\href{https://doi.org/10.5281/zenodo.23791922}{\texttt{10.5281/zenodo.23791922}}}
\newcommand{\dataDOI}{\href{https://doi.org/10.5281/zenodo.23791585}{\texttt{10.5281/zenodo.23791585}}}

\title{Testing hardware quantum machine learning claims\\ with a null circuit: fill prediction on Binance futures\\ with 129 qubits of an IBM Heron processor}
\author{Muhammad Faryad\\[2pt]
\small Department of Physics, Syed Babar Ali School of Science and Engineering,\\
\small Lahore University of Management Sciences (LUMS), Lahore 54792, Pakistan\\
\small \href{mailto:muhammadfaryad@gmail.com}{muhammadfaryad@gmail.com}}
\date{\small \today}

\begin{document}
\maketitle

\begin{abstract}
Quantum processors are now used as feature generators for classical models on tabular data. A circuit encodes each
record, the device returns expectation values, and a classical model learns from them. A recent report claimed
large gains from such features, gains that appeared only on noisy hardware. These claims are hard to
check, because a quantum device is not a static function. Its state changes from job to job and from one calibration
to the next, and that drift can leak information about the outcome into the features. We propose a five-step protocol
for testing claims of this kind. It runs a circuit that carries no data in the same jobs as the real one. It checks
whether the hardware leaves a recognizable mark on each job. It compares random and natural assignments of records to
jobs, replays the null results under other submission orders, and closes the leak with a job-purged backtest. We apply
it, pre-registered, to the HSBC and IBM study of fill probabilities in bond trading, which reported a large hardware-only
gain. We use a public analog task and run the study's circuit on a 129-qubit chain of an IBM Heron processor. Under
controlled assignment the hardware features give no gain over classical features, and the device reproduces the ideal
features about as well as shot noise allows. Yet every job carries a distinct fingerprint. A circuit with no data at
all predicts fills well above chance when records are batched by outcome and the same job sits on both sides of the
train/test split. Removing that overlap removes the effect. The protocol costs under an hour of processor time and is designed for any hardware feature or kernel pipeline on tabular data.
\end{abstract}

%=====================================================================================================================
\section{Introduction}

A simple way to put a quantum computer to work in machine learning is to use it as a feature map
\cite{havlicek2019,schuld2019,huang2021}. A parameterized circuit encodes each record. The device returns a set of
expectation values. A classical model is then trained on those values. The learning stays classical, and the quantum
part is a fixed transform of the input. The idea is attractive because it needs no training on the device and fits
existing data pipelines.

It also raises a measurement problem that is easy to miss. A simulator is a deterministic function: the same input
gives the same output. A quantum processor is not. Its gate and readout
errors change between calibrations, and they also drift from one job to the next. Each job therefore stamps its own small offset on every feature it produces. If
the way records are grouped into jobs is related to the outcome, the model can learn the stamp instead of the record.
This is the batch effect that has misled genomics for years \cite{leek2010}, and a form of the leakage that affects
applied machine learning broadly \cite{kaufman2012,kapoor2023}.

The question became concrete in September 2025. A study by HSBC and IBM used features measured on IBM Heron
processors to predict whether quotes on European corporate bonds would be filled \cite{ciceri2025}. The hardware
features raised the out-of-sample AUC far above what classical features gave. The same circuit in noiseless simulation
gave no gain at all. The authors reported this as an observation they could not explain. Critics argued that a gain
which disappears in simulation is not a quantum computational effect \cite{aaronson2025,dunbar2025}, and commenters
raised overfitting, look-ahead and leakage as possible causes \cite{aaronson2025}. A later study of projected feature maps on biomedical data found
gains comparable to earlier reports only when samples were run in class order, and traced part of them to
order-dependent hardware drift \cite{docarmo2026}.

This paper does two things. First, it sets out a protocol for testing hardware feature claims of this kind
(Sec.~\ref{sec:protocol}). The protocol is built around a \emph{null circuit}: the same physical circuit with every
data angle set to zero, run in the same jobs as the data. Its output cannot depend on the record, so whatever it
``predicts'' comes from the device. Second, it applies the protocol to the HSBC/IBM claim, pre-registered, on 129
qubits of \texttt{ibm\_fez} (Secs.~\ref{sec:study} and \ref{sec:results}). We cannot use HSBC's data, which are
private. We build a public task with the same structure and a similar classical baseline, and we fix the design,
predictions and decision rules before the first run.

The answer has two parts. Under controlled assignment there is no gain. But the device does leave a fingerprint on
every job, strong enough that a circuit with no data predicts fills well above chance under one specific pipeline.
The protocol identifies that pipeline and shows how to close it.

%=====================================================================================================================
\section{The claim under test}
\label{sec:claim}

Table~\ref{tab:hsbc} summarizes the HSBC/IBM study \cite{ciceri2025} as far as it reports its setup. The circuit is a
Trotterized Heisenberg evolution on a chain of qubits, with couplings set by the features. Two settings were run.
The ``longer'' setting, with two blocks and a small evolution time, carries the largest gain. The AUC values are
medians over rolling test windows.

\begin{table}[h]
\centering\small
\begin{tabularx}{\linewidth}{@{}>{\raggedright\arraybackslash}p{3.3cm}X@{}}
\toprule
Data & HSBC requests for quote on European corporate bonds; 16{,}000 events processed on hardware; label = quote filled (about 37\%) \\
Features & 216 per event, standardized, then mapped to $J = 2\pi\tanh(z/3)$ \\
Circuit & Random product state, then $R_{ZZ}$, $R_{YY}$, $R_{XX}$ on nearest neighbors of a 109-qubit chain; ``shorter'': 1 block, $\alpha=1$; ``longer'': 2 blocks, $\alpha=0.1$ \\
Readout & $\langle X\rangle$, $\langle Y\rangle$, $\langle Z\rangle$ on every qubit; 4{,}096 shots; Pauli twirling and TREX \cite{wallman2016,vandenberg2022} \\
Devices & \texttt{ibm\_torino}; checks on \texttt{ibm\_fez} and \texttt{ibm\_marrakesh} \\
Protocol & Rolling backtest; training sets of 500 to 2{,}000 past events; a blinding window between training and test; four model families tuned by grid search \\
Reported AUC (no blinding) & classical about 0.63 (averaged over blinding windows); noiseless simulation 0.60; hardware ``shorter'' 0.75; hardware ``longer'' 0.97 \\
Reported AUC (1-day blinding) & classical about 0.63; hardware ``longer'' 0.88, decaying roughly linearly up to 4 days \\
\bottomrule
\end{tabularx}
\caption{The study under test \cite{ciceri2025}.}
\label{tab:hsbc}
\end{table}

The public objections fall into a few groups. The gain vanishes in noiseless simulation, so it cannot be a quantum
computational effect \cite{aaronson2025,dunbar2025}. The mechanism is unexplained, as the authors themselves note.
Commenters on Ref.~\cite{aaronson2025} added that the gain shrinks as the blinding window grows, which points to
leakage or look-ahead; that trial counts and significance tests were not reported; and that the noiseless reference was
a matrix-product-state simulation rather than an exact one. Ivezic argued, by contrast, that the authors never claimed
a quantum advantage \cite{ivezic2025}, and The Quantum Insider gave a longer account of the study \cite{tqi2025}. And
the data are private, so nobody can check the result directly. A useful test has to answer as many of these as it can with data.

%=====================================================================================================================
\section{How a hardware feature pipeline can leak}
\label{sec:leak}

\begin{figure}[t]
\centering
\resizebox{\linewidth}{!}{\begin{tikzpicture}[x=1cm,y=1cm]
\node[anchor=east,font=\small\bfseries] at (-0.25,0) {(a)};
\fill[cY!75,draw=cY!60!black,rounded corners=1pt] (0.00,-0.22) rectangle ++(0.5,0.44);
\fill[white] (0.25,0) circle (0.09);
\fill[cT!75,draw=cT!60!black,rounded corners=1pt] (0.62,-0.22) rectangle ++(0.5,0.44);
\fill[cY!75,draw=cY!60!black,rounded corners=1pt] (1.24,-0.22) rectangle ++(0.5,0.44);
\fill[cT!75,draw=cT!60!black,rounded corners=1pt] (1.86,-0.22) rectangle ++(0.5,0.44);
\fill[white] (2.11,0) circle (0.09);
\fill[cY!75,draw=cY!60!black,rounded corners=1pt] (2.48,-0.22) rectangle ++(0.5,0.44);
\fill[cT!75,draw=cT!60!black,rounded corners=1pt] (3.10,-0.22) rectangle ++(0.5,0.44);
\fill[white] (3.35,0) circle (0.09);
\fill[cC!75,draw=cC!60!black,rounded corners=1pt] (3.72,-0.22) rectangle ++(0.5,0.44);
\fill[cM!75,draw=cM!60!black,rounded corners=1pt] (4.34,-0.22) rectangle ++(0.5,0.44);
\fill[white] (4.59,0) circle (0.09);
\fill[cC!75,draw=cC!60!black,rounded corners=1pt] (4.96,-0.22) rectangle ++(0.5,0.44);
\fill[white] (5.21,0) circle (0.09);
\fill[cM!75,draw=cM!60!black,rounded corners=1pt] (5.58,-0.22) rectangle ++(0.5,0.44);
\fill[cC!75,draw=cC!60!black,rounded corners=1pt] (6.20,-0.22) rectangle ++(0.5,0.44);
\fill[cM!75,draw=cM!60!black,rounded corners=1pt] (6.82,-0.22) rectangle ++(0.5,0.44);
\fill[white] (7.07,0) circle (0.09);
\fill[cC!75,draw=cC!60!black,rounded corners=1pt] (7.44,-0.22) rectangle ++(0.5,0.44);
\fill[cM!75,draw=cM!60!black,rounded corners=1pt] (8.06,-0.22) rectangle ++(0.5,0.44);
\fill[white] (8.31,0) circle (0.09);
\fill[cC!75,draw=cC!60!black,rounded corners=1pt] (8.68,-0.22) rectangle ++(0.5,0.44);
\fill[white] (8.93,0) circle (0.09);
\fill[cM!75,draw=cM!60!black,rounded corners=1pt] (9.30,-0.22) rectangle ++(0.5,0.44);
\draw[decorate,decoration={brace,amplitude=3pt,mirror},thick,cG] (0,-0.32) -- (6.70,-0.32) node[midway,below=2pt,font=\scriptsize,cG] {training window};
\draw[decorate,decoration={brace,amplitude=3pt,mirror},thick,cR] (6.82,-0.32) -- (9.80,-0.32) node[midway,below=2pt,font=\scriptsize,cR] {test block};
\node[anchor=west,font=\footnotesize,text width=5.4cm] at (10.4,0) {Each job holds both outcomes. Its offset tells the model nothing. \textcolor{cN}{\textbf{No leak.}}};
\node[anchor=east,font=\small\bfseries] at (-0.25,-2.0) {(b)};
\fill[cT!75,draw=cT!60!black,rounded corners=1pt] (0.00,-2.22) rectangle ++(0.5,0.44);
\fill[white] (0.25,-2.0) circle (0.09);
\fill[cY!75,draw=cY!60!black,rounded corners=1pt] (0.62,-2.22) rectangle ++(0.5,0.44);
\fill[cY!75,draw=cY!60!black,rounded corners=1pt] (1.24,-2.22) rectangle ++(0.5,0.44);
\fill[cT!75,draw=cT!60!black,rounded corners=1pt] (1.86,-2.22) rectangle ++(0.5,0.44);
\fill[white] (2.11,-2.0) circle (0.09);
\fill[cY!75,draw=cY!60!black,rounded corners=1pt] (2.48,-2.22) rectangle ++(0.5,0.44);
\fill[cT!75,draw=cT!60!black,rounded corners=1pt] (3.10,-2.22) rectangle ++(0.5,0.44);
\fill[white] (3.35,-2.0) circle (0.09);
\fill[cC!75,draw=cC!60!black,rounded corners=1pt] (3.72,-2.22) rectangle ++(0.5,0.44);
\fill[cM!75,draw=cM!60!black,rounded corners=1pt] (4.34,-2.22) rectangle ++(0.5,0.44);
\fill[white] (4.59,-2.0) circle (0.09);
\fill[cM!75,draw=cM!60!black,rounded corners=1pt] (4.96,-2.22) rectangle ++(0.5,0.44);
\fill[white] (5.21,-2.0) circle (0.09);
\fill[cC!75,draw=cC!60!black,rounded corners=1pt] (5.58,-2.22) rectangle ++(0.5,0.44);
\fill[cC!75,draw=cC!60!black,rounded corners=1pt] (6.20,-2.22) rectangle ++(0.5,0.44);
\fill[cM!75,draw=cM!60!black,rounded corners=1pt] (6.82,-2.22) rectangle ++(0.5,0.44);
\fill[white] (7.07,-2.0) circle (0.09);
\fill[cC!75,draw=cC!60!black,rounded corners=1pt] (7.44,-2.22) rectangle ++(0.5,0.44);
\fill[cM!75,draw=cM!60!black,rounded corners=1pt] (8.06,-2.22) rectangle ++(0.5,0.44);
\fill[white] (8.31,-2.0) circle (0.09);
\fill[cM!75,draw=cM!60!black,rounded corners=1pt] (8.68,-2.22) rectangle ++(0.5,0.44);
\fill[white] (8.93,-2.0) circle (0.09);
\fill[cC!75,draw=cC!60!black,rounded corners=1pt] (9.30,-2.22) rectangle ++(0.5,0.44);
\draw[decorate,decoration={brace,amplitude=3pt,mirror},thick,cG] (0,-2.32) -- (6.70,-2.32) node[midway,below=2pt,font=\scriptsize,cG] {training window};
\draw[decorate,decoration={brace,amplitude=3pt,mirror},thick,cR] (6.82,-2.32) -- (9.80,-2.32) node[midway,below=2pt,font=\scriptsize,cR] {test block};
\node[anchor=west,font=\footnotesize,text width=5.4cm] at (10.4,-2.0) {Fills share a job. The model learns the \textcolor{cM}{magenta} job's offset in training and finds it again in the test block. \textcolor{cR}{\textbf{Leak.}}};
\node[anchor=east,font=\small\bfseries] at (-0.25,-4.0) {(c)};
\fill[cT!75,draw=cT!60!black,rounded corners=1pt] (0.00,-4.22) rectangle ++(0.5,0.44);
\fill[white] (0.25,-4.0) circle (0.09);
\fill[cY!75,draw=cY!60!black,rounded corners=1pt] (0.62,-4.22) rectangle ++(0.5,0.44);
\fill[cY!75,draw=cY!60!black,rounded corners=1pt] (1.24,-4.22) rectangle ++(0.5,0.44);
\fill[cT!75,draw=cT!60!black,rounded corners=1pt] (1.86,-4.22) rectangle ++(0.5,0.44);
\fill[white] (2.11,-4.0) circle (0.09);
\fill[cY!75,draw=cY!60!black,rounded corners=1pt] (2.48,-4.22) rectangle ++(0.5,0.44);
\fill[cT!75,draw=cT!60!black,rounded corners=1pt] (3.10,-4.22) rectangle ++(0.5,0.44);
\fill[white] (3.35,-4.0) circle (0.09);
\fill[gray!25,draw=gray!60,rounded corners=1pt] (3.72,-4.22) rectangle ++(0.5,0.44);
\draw[gray!70,line width=0.6pt] (3.77,-4.17) -- (4.17,-3.83);
\fill[gray!25,draw=gray!60,rounded corners=1pt] (4.34,-4.22) rectangle ++(0.5,0.44);
\fill[gray!60] (4.59,-4.0) circle (0.09);
\draw[gray!70,line width=0.6pt] (4.39,-4.17) -- (4.79,-3.83);
\fill[gray!25,draw=gray!60,rounded corners=1pt] (4.96,-4.22) rectangle ++(0.5,0.44);
\fill[gray!60] (5.21,-4.0) circle (0.09);
\draw[gray!70,line width=0.6pt] (5.01,-4.17) -- (5.41,-3.83);
\fill[gray!25,draw=gray!60,rounded corners=1pt] (5.58,-4.22) rectangle ++(0.5,0.44);
\draw[gray!70,line width=0.6pt] (5.63,-4.17) -- (6.03,-3.83);
\fill[gray!25,draw=gray!60,rounded corners=1pt] (6.20,-4.22) rectangle ++(0.5,0.44);
\draw[gray!70,line width=0.6pt] (6.25,-4.17) -- (6.65,-3.83);
\fill[cM!75,draw=cM!60!black,rounded corners=1pt] (6.82,-4.22) rectangle ++(0.5,0.44);
\fill[white] (7.07,-4.0) circle (0.09);
\fill[cC!75,draw=cC!60!black,rounded corners=1pt] (7.44,-4.22) rectangle ++(0.5,0.44);
\fill[cM!75,draw=cM!60!black,rounded corners=1pt] (8.06,-4.22) rectangle ++(0.5,0.44);
\fill[white] (8.31,-4.0) circle (0.09);
\fill[cM!75,draw=cM!60!black,rounded corners=1pt] (8.68,-4.22) rectangle ++(0.5,0.44);
\fill[white] (8.93,-4.0) circle (0.09);
\fill[cC!75,draw=cC!60!black,rounded corners=1pt] (9.30,-4.22) rectangle ++(0.5,0.44);
\draw[decorate,decoration={brace,amplitude=3pt,mirror},thick,cG] (0,-4.32) -- (6.70,-4.32) node[midway,below=2pt,font=\scriptsize,cG] {training window};
\draw[decorate,decoration={brace,amplitude=3pt,mirror},thick,cR] (6.82,-4.32) -- (9.80,-4.32) node[midway,below=2pt,font=\scriptsize,cR] {test block};
\node[anchor=west,font=\footnotesize,text width=5.4cm] at (10.4,-4.0) {Job-purged backtest: training rows that share a job with the test block are dropped. \textcolor{cN}{\textbf{No leak.}}};
\draw[-{Stealth[length=2.5mm]},thick,cG] (0,0.55) -- (9.85,0.55) node[above left,font=\scriptsize] {time};
\begin{scope}[shift={(0.2,-5.3)},font=\scriptsize]
\fill[cY!75,draw=cY!60!black] (0.0,0) rectangle ++(0.35,0.3); \node[anchor=west] at (0.38,0.15) {job 1a};
\fill[cT!75,draw=cT!60!black] (1.55,0) rectangle ++(0.35,0.3); \node[anchor=west] at (1.9300000000000002,0.15) {job 1b};
\fill[cC!75,draw=cC!60!black] (3.1,0) rectangle ++(0.35,0.3); \node[anchor=west] at (3.48,0.15) {job 2a};
\fill[cM!75,draw=cM!60!black] (4.65,0) rectangle ++(0.35,0.3); \node[anchor=west] at (5.03,0.15) {job 2b};
\fill[cG!60] (6.5,0.15) circle (0.09); \node[anchor=west] at (6.6,0.15) {fill (label 1)};
\end{scope}
\end{tikzpicture}}
\caption{How a batch leak works. Each square is a record, colored by the hardware job that produced it; a dot marks a
fill. (a)~Records batched at random or in time order: every job holds both outcomes, so the job's offset carries no
information. (b)~Records batched by outcome, with the same job on both sides of the train/test split: the model learns
the job's offset from the training rows and ``predicts'' the test rows. (c)~A job-purged backtest drops training rows
that share a job with the test block, and the leak disappears.}
\label{fig:mechanism}
\end{figure}
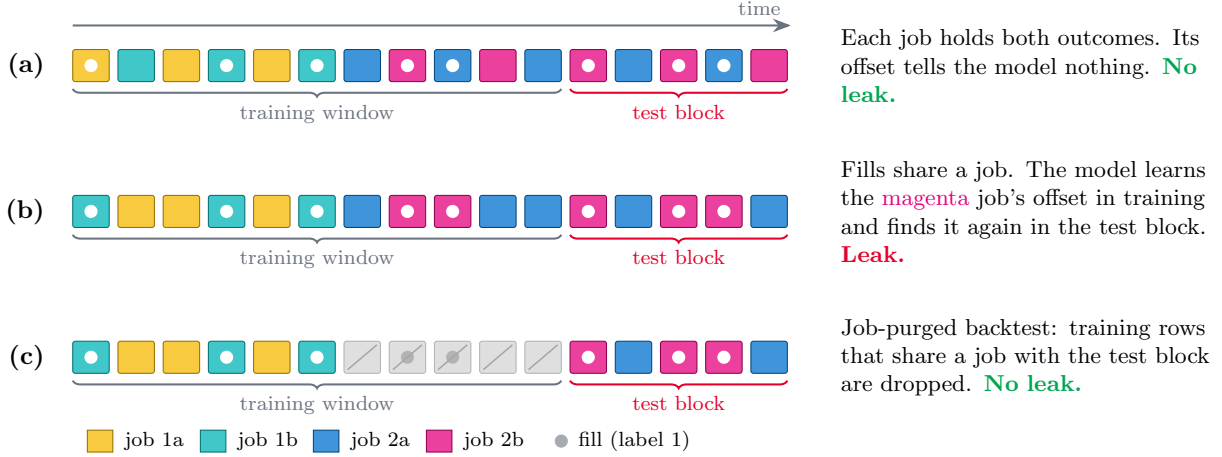

A pure time stamp cannot leak in a forward rolling backtest. Every test record comes after its training records, and
on our task a model that sees only the time does no better than chance with no blinding window. What can leak is the batch. Suppose each
hardware job adds a small, fixed offset to every feature it produces. On its own this is harmless noise. It becomes a
signal when two things happen together (Fig.~\ref{fig:mechanism}). The records in a job share an outcome, for example
because they were submitted in class order. And some records of the same job sit in the training window while others
sit in the test block. The model then learns ``this offset means a fill'' from the training rows and applies it to the
test rows. Hundreds of features, each with a tiny offset, add up to a clear signature.

This is why the blinding window matters. A rolling backtest with no blinding places test records next to the
training records they were processed with. Widening the window separates them, and a leak of this kind should fade.
That is the pattern HSBC reported.

%=====================================================================================================================
\section{The protocol}
\label{sec:protocol}

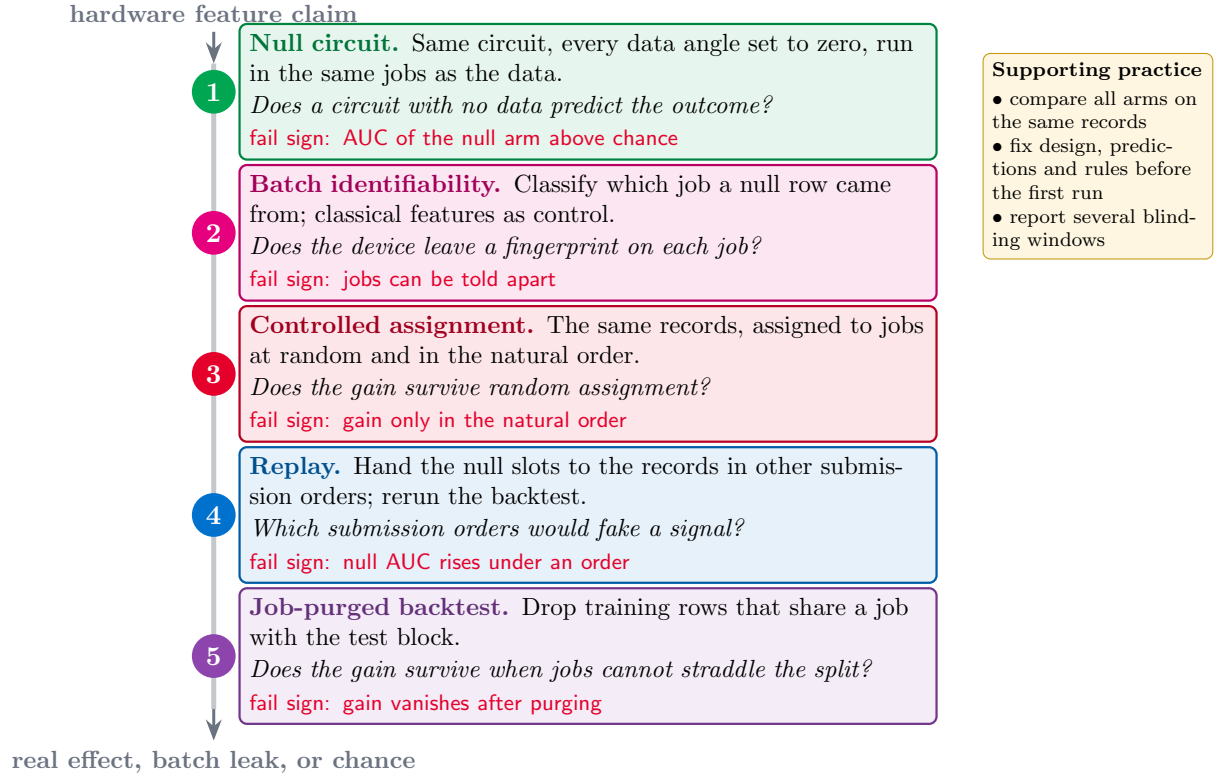
\begin{figure}[t]
\centering
\resizebox{0.97\linewidth}{!}{\begin{tikzpicture}[font=\small,box/.style={rounded corners=3pt,draw=#1!80!black,fill=#1!10,line width=0.9pt,minimum height=1.5cm,text width=9.6cm,inner sep=4pt},num/.style={circle,fill=#1,text=white,font=\bfseries,minimum size=0.62cm,inner sep=0pt}]
\node[box=cN,anchor=west] (b1) at (0.55,0.0) {\textbf{\textcolor{cN!70!black}{Null circuit.}} Same circuit, every data angle set to zero, run in the same jobs as the data.\\[1pt]\emph{Does a circuit with no data predict the outcome?}\\[1pt]\textcolor{cR}{\footnotesize\textsf{fail sign: AUC of the null arm above chance}}};
\node[num=cN] at (0.2,0.0) {1};
\node[box=cM,anchor=west] (b2) at (0.55,-2.0) {\textbf{\textcolor{cM!70!black}{Batch identifiability.}} Classify which job a null row came from; classical features as control.\\[1pt]\emph{Does the device leave a fingerprint on each job?}\\[1pt]\textcolor{cR}{\footnotesize\textsf{fail sign: jobs can be told apart}}};
\node[num=cM] at (0.2,-2.0) {2};
\node[box=cR,anchor=west] (b3) at (0.55,-4.0) {\textbf{\textcolor{cR!70!black}{Controlled assignment.}} The same records, assigned to jobs at random and in the natural order.\\[1pt]\emph{Does the gain survive random assignment?}\\[1pt]\textcolor{cR}{\footnotesize\textsf{fail sign: gain only in the natural order}}};
\node[num=cR] at (0.2,-4.0) {3};
\node[box=cC,anchor=west] (b4) at (0.55,-6.0) {\textbf{\textcolor{cC!70!black}{Replay.}} Hand the null slots to the records in other submission orders; rerun the backtest.\\[1pt]\emph{Which submission orders would fake a signal?}\\[1pt]\textcolor{cR}{\footnotesize\textsf{fail sign: null AUC rises under an order}}};
\node[num=cC] at (0.2,-6.0) {4};
\node[box=cQ,anchor=west] (b5) at (0.55,-8.0) {\textbf{\textcolor{cQ!70!black}{Job-purged backtest.}} Drop training rows that share a job with the test block.\\[1pt]\emph{Does the gain survive when jobs cannot straddle the split?}\\[1pt]\textcolor{cR}{\footnotesize\textsf{fail sign: gain vanishes after purging}}};
\node[num=cQ] at (0.2,-8.0) {5};
\draw[-{Stealth[length=3mm]},line width=1.2pt,cG] (0.2,0.85) -- (0.2,0.4);
\node[font=\small\bfseries,cG,anchor=south] at (0.2,0.85) {hardware feature claim};
\draw[-{Stealth[length=3mm]},line width=1.2pt,cG] (0.2,-8.75) -- (0.2,-9.2) node[below,font=\small\bfseries,cG] {real effect, batch leak, or chance};
\begin{scope}[on background layer]\draw[cG!40,line width=2pt] (0.2,0.4) -- (0.2,-8.75);\end{scope}
\node[draw=cY!80!black,fill=cY!12,rounded corners=3pt,font=\scriptsize,text width=3.0cm,align=left,anchor=north west] at (11.1,0.55) {\textbf{Supporting practice}\\[2pt]\textbullet~compare all arms on the same records\\\textbullet~fix design, predictions and rules before the first run\\\textbullet~report several blinding windows};
\end{tikzpicture}}
\caption{The five-step protocol. Each step asks one question and has a clear failure sign. Steps 1 and 3 cost
processor time, and run in the same jobs as the data. Steps 2, 4 and 5 are analyses of the data already taken.}
\label{fig:protocol}
\end{figure}

The protocol has five steps (Fig.~\ref{fig:protocol}). Each one is cheap, and each one can fail on its own. Together
they separate a real hardware effect from a batch leak and from chance.

\paragraph{Step 1: a null circuit in the same jobs.} Run the data circuit with every data angle set to zero. It
executes the same physical gates and sees the same noise, but its output cannot depend on the record. Interleave its
rows with the data rows in every job, so both see identical hardware conditions. If the null circuit ``predicts'' the
outcome under the study's own pipeline, the pipeline leaks.

\paragraph{Step 2: batch identifiability.} Train a classifier to tell which job a null row came from. If it can, the
device leaves a fingerprint, and any pipeline that groups records by outcome can leak. A control with the classical
features of the same rows must stay at chance.

\paragraph{Step 3: controlled assignment.} Process the same records two ways, in the same jobs: assigned to jobs at
random, and in the study's natural order. A real hardware effect survives random assignment. A leak does not.

\paragraph{Step 4: replay.} Null rows do not depend on the record. So the null slots of each job can be handed to the
records in any order after the fact, and the backtest rerun. This measures, with no extra processor time, how much
apparent signal each submission order would produce on this device.

\paragraph{Step 5: a job-purged backtest.} In each training window, drop every row that shares a job with the test
block. A leak of the kind in Fig.~\ref{fig:mechanism} disappears. A real effect does not.

Two practices support the steps. Fix the design, the predictions and the decision rules before the first run, so the
selective-reporting objection has a direct answer \cite{nosek2018}. And compare every arm on the same records, because
a single backtest AUC varies a lot between record sets.

%=====================================================================================================================
\section{Applying the protocol on 129 qubits}
\label{sec:study}

\subsection{A public analog task}

We build the task from public Binance USD-M perpetual futures data \cite{binance}: trades, order-book depth and
positioning metrics for eight liquid contracts over eight days in September 2026. An event is a passive limit order
placed at a random time. Its label is 1 if a trade touches the order's price within 60 seconds. This ``touch'' rule ignores
queue position, so it is an optimistic proxy for a fill; studies with order-level data can follow real executions
instead \cite{arroyo2024}. The quote distance is set so that about 37\% of orders fill, as in
HSBC's data. Each event has 216 features of the order's own contract, the market leaders, the cross-section, the order
book and positioning. Table~\ref{tab:task} compares the two tasks.

\begin{table}[h]
\centering\small
\begin{tabularx}{\linewidth}{@{}>{\raggedright\arraybackslash}p{3.2cm}>{\raggedright\arraybackslash}X>{\raggedright\arraybackslash}X@{}}
\toprule
 & HSBC/IBM \cite{ciceri2025} & This study \\
\midrule
Event & request for quote on a bond & passive limit order on a crypto future \\
Label & quote filled & order touched within 60\,s \\
Fill rate & about 37\% & 34\% on the hardware events \\
Features & 216 & 216 \\
Events on hardware & 16{,}000 & 1{,}000 (125 per day, 8 days) \\
Classical AUC & 0.63 & 0.58 on the hardware events \\
\bottomrule
\end{tabularx}
\caption{The public analog task. Our AUC uses the same rolling backtest as all other results (Sec.~\ref{sec:analysis}).}
\label{tab:task}
\end{table}

A single backtest AUC is a noisy number. On five other random 1{,}000-event subsets of our pool, the classical AUC ranges
from 0.51 to 0.64. This is why every comparison below is paired on the same 1{,}000 events.

\subsection{Circuits}

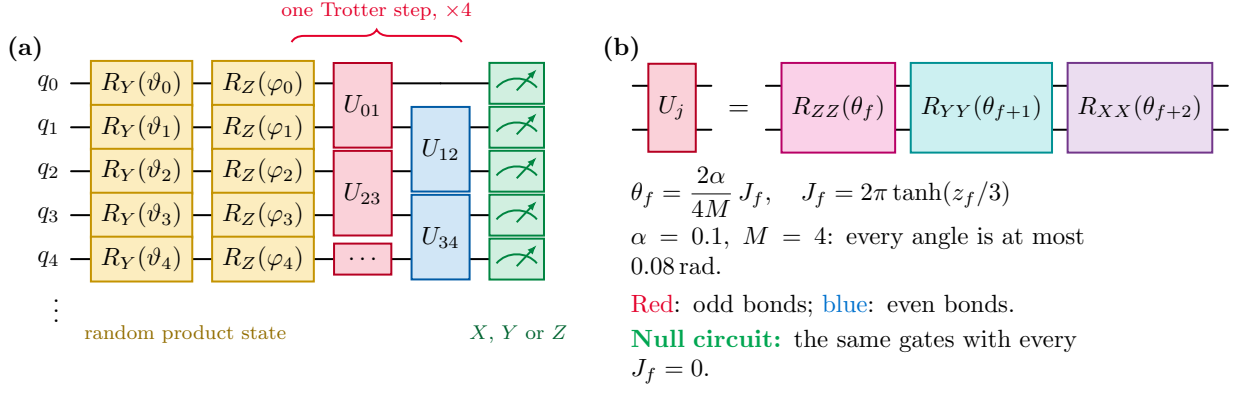
\begin{figure}[t]
\centering
\resizebox{\linewidth}{!}{\begin{tikzpicture}
\node[anchor=north west] (A) at (0,0) {\begin{quantikz}[row sep={0.62cm,between origins},column sep=0.28cm,font=\small]
\lstick{$q_0$} & \gate[style={fill=cY!25,draw=cY!80!black}]{R_Y(\vartheta_0)} & \gate[style={fill=cY!25,draw=cY!80!black}]{R_Z(\varphi_0)} & \gate[2,style={fill=cR!18,draw=cR!80!black}]{U_{01}} & & \meter[style={fill=cN!18,draw=cN!80!black}]{} \\
\lstick{$q_1$} & \gate[style={fill=cY!25,draw=cY!80!black}]{R_Y(\vartheta_1)} & \gate[style={fill=cY!25,draw=cY!80!black}]{R_Z(\varphi_1)} & & \gate[2,style={fill=cC!18,draw=cC!80!black}]{U_{12}} & \meter[style={fill=cN!18,draw=cN!80!black}]{} \\
\lstick{$q_2$} & \gate[style={fill=cY!25,draw=cY!80!black}]{R_Y(\vartheta_2)} & \gate[style={fill=cY!25,draw=cY!80!black}]{R_Z(\varphi_2)} & \gate[2,style={fill=cR!18,draw=cR!80!black}]{U_{23}} & & \meter[style={fill=cN!18,draw=cN!80!black}]{} \\
\lstick{$q_3$} & \gate[style={fill=cY!25,draw=cY!80!black}]{R_Y(\vartheta_3)} & \gate[style={fill=cY!25,draw=cY!80!black}]{R_Z(\varphi_3)} & & \gate[2,style={fill=cC!18,draw=cC!80!black}]{U_{34}} & \meter[style={fill=cN!18,draw=cN!80!black}]{} \\
\lstick{$q_4$} & \gate[style={fill=cY!25,draw=cY!80!black}]{R_Y(\vartheta_4)} & \gate[style={fill=cY!25,draw=cY!80!black}]{R_Z(\varphi_4)} & \gate[style={fill=cR!18,draw=cR!80!black}]{\,\cdots\,} & & \meter[style={fill=cN!18,draw=cN!80!black}]{} \\
\lstick{$\vdots$} \setwiretype{n} & & & & & 
\end{quantikz}};
\node[anchor=north west,font=\small\bfseries] at (-0.3,0.1) {(a)};
\draw[decorate,decoration={brace,amplitude=5pt},thick,cR] ($(A.north west)+(3.85,-0.12)$) -- ($(A.north west)+(6.25,-0.12)$) node[midway,above=5pt,font=\scriptsize,cR] {one Trotter step, $\times 4$};
\node[font=\scriptsize,cY!60!black,anchor=north] at ($(A.south west)+(2.35,0.35)$) {random product state};
\node[font=\scriptsize,cN!60!black,anchor=north] at ($(A.south west)+(7.05,0.35)$) {$X$, $Y$ or $Z$};
\node[anchor=north west,font=\small\bfseries] at (8.1,0.1) {(b)};
\node[anchor=north west] (B) at (8.4,-0.15) {\begin{quantikz}[row sep={0.62cm,between origins},column sep=0.22cm,font=\small]
& \gate[2,style={fill=cR!18,draw=cR!80!black}]{U_j} & \\
& & 
\end{quantikz}\;=\;\begin{quantikz}[row sep={0.62cm,between origins},column sep=0.22cm,font=\small]
& \gate[2,style={fill=cM!18,draw=cM!80!black}]{R_{ZZ}(\theta_{f})} & \gate[2,style={fill=cT!20,draw=cT!80!black}]{R_{YY}(\theta_{f+1})} & \gate[2,style={fill=cQ!18,draw=cQ!80!black}]{R_{XX}(\theta_{f+2})} & \\
& & & & 
\end{quantikz}};
\node[anchor=north west,font=\small,text width=7.0cm] at (8.5,-1.75) {$\theta_f=\dfrac{2\alpha}{4M}\,J_f,\quad J_f=2\pi\tanh(z_f/3)$\\[3pt]$\alpha=0.1,\ M=4$: every angle is at most $0.08$\,rad.\\[3pt]\textcolor{cR}{Red}: odd bonds; \textcolor{cC}{blue}: even bonds.\\[2pt]\textcolor{cN}{\textbf{Null circuit:}} the same gates with every $J_f=0$.};
\end{tikzpicture}}
\caption{The circuit. (a)~A random product state, then four Trotter steps of two-qubit blocks on the odd and the even
bonds of a 129-qubit chain, then a readout of every qubit in the $X$, $Y$ or $Z$ basis. Five of the 129 qubits are
shown. (b)~Each block applies $R_{ZZ}$, $R_{YY}$ and $R_{XX}$ with angles set by three features. The null circuit is
the same transpiled circuit with every $J_f$ set to zero.}
\label{fig:circuit}
\end{figure}

We implement HSBC's circuit family with a stated assignment of features to gates (Fig.~\ref{fig:circuit}). The
``longer'' circuit, which carries the largest reported gain, runs on a 129-qubit chain of \texttt{ibm\_fez}, the
longest simple path we found on its heavy-hex lattice. It has 1{,}536 two-qubit rotations, which become 3{,}072 CZ gates
in 48 two-qubit layers. Every data-dependent angle is small, at most 0.08\,rad, so its ideal output depends only weakly
on the data. Across events, its features vary about ten times less than those of the ``shorter'' circuit. Small angles
also make it easy to simulate exactly. Pauli propagation \cite{rudolph2025} gives all 387 features for 1{,}000 events on
a laptop and agrees with an exact light-cone statevector to $3\times10^{-6}$. Our noiseless reference is therefore
exact, which answers the objection about an approximate reference.

\subsection{Arms and schedule}

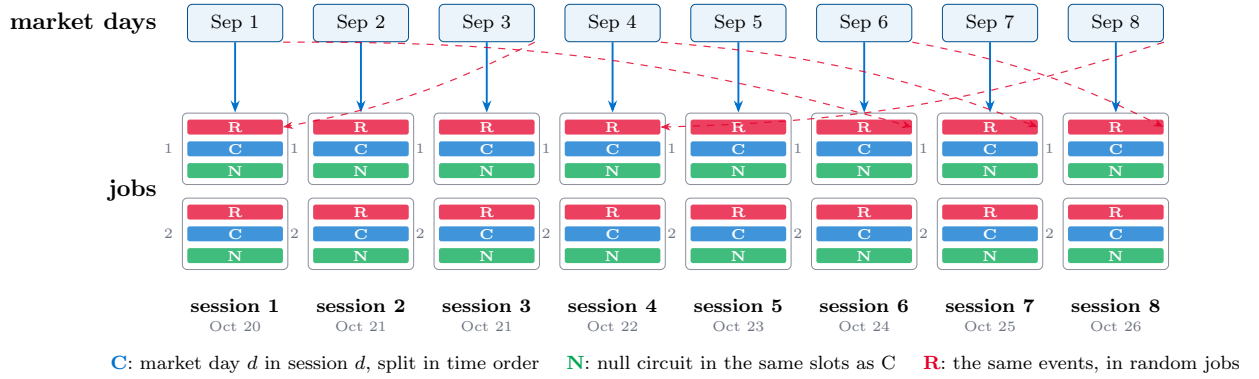
\begin{figure}[t]
\centering
\resizebox{\linewidth}{!}{\begin{tikzpicture}[font=\scriptsize,x=1.85cm,y=1cm]
\node[draw=cC!70!black,fill=cC!8,rounded corners=2pt,minimum width=1.4cm,minimum height=0.55cm] (m0) at (0,2.6) {Sep 1};
\node[draw=cC!70!black,fill=cC!8,rounded corners=2pt,minimum width=1.4cm,minimum height=0.55cm] (m1) at (1,2.6) {Sep 2};
\node[draw=cC!70!black,fill=cC!8,rounded corners=2pt,minimum width=1.4cm,minimum height=0.55cm] (m2) at (2,2.6) {Sep 3};
\node[draw=cC!70!black,fill=cC!8,rounded corners=2pt,minimum width=1.4cm,minimum height=0.55cm] (m3) at (3,2.6) {Sep 4};
\node[draw=cC!70!black,fill=cC!8,rounded corners=2pt,minimum width=1.4cm,minimum height=0.55cm] (m4) at (4,2.6) {Sep 5};
\node[draw=cC!70!black,fill=cC!8,rounded corners=2pt,minimum width=1.4cm,minimum height=0.55cm] (m5) at (5,2.6) {Sep 6};
\node[draw=cC!70!black,fill=cC!8,rounded corners=2pt,minimum width=1.4cm,minimum height=0.55cm] (m6) at (6,2.6) {Sep 7};
\node[draw=cC!70!black,fill=cC!8,rounded corners=2pt,minimum width=1.4cm,minimum height=0.55cm] (m7) at (7,2.6) {Sep 8};
\node[anchor=east,font=\small\bfseries] at (-0.55,2.6) {market days};
\node[anchor=east,font=\small\bfseries] at (-0.55,0.15) {jobs};
\node[draw=cG!80,fill=white,rounded corners=2pt,minimum width=1.55cm,minimum height=1.05cm] (j00) at (0,0.75) {};
\fill[cR!75,rounded corners=1pt] (-0.38,0.9600000000000001) rectangle (0.38,1.1800000000000002);
\node[font=\tiny\bfseries,white] at (0,1.07) {R};
\fill[cC!75,rounded corners=1pt] (-0.38,0.64) rectangle (0.38,0.86);
\node[font=\tiny\bfseries,white] at (0,0.75) {C};
\fill[cN!75,rounded corners=1pt] (-0.38,0.32000000000000006) rectangle (0.38,0.54);
\node[font=\tiny\bfseries,white] at (0,0.43000000000000005) {N};
\node[font=\tiny,cG,anchor=east] at (-0.42,0.75) {1};
\node[draw=cG!80,fill=white,rounded corners=2pt,minimum width=1.55cm,minimum height=1.05cm] (j01) at (0,-0.5) {};
\fill[cR!75,rounded corners=1pt] (-0.38,-0.29) rectangle (0.38,-0.06999999999999999);
\node[font=\tiny\bfseries,white] at (0,-0.18) {R};
\fill[cC!75,rounded corners=1pt] (-0.38,-0.61) rectangle (0.38,-0.39);
\node[font=\tiny\bfseries,white] at (0,-0.5) {C};
\fill[cN!75,rounded corners=1pt] (-0.38,-0.93) rectangle (0.38,-0.7100000000000001);
\node[font=\tiny\bfseries,white] at (0,-0.8200000000000001) {N};
\node[font=\tiny,cG,anchor=east] at (-0.42,-0.5) {2};
\node[font=\scriptsize\bfseries] at (0,-1.55) {session 1};
\node[font=\tiny,cG] at (0,-1.85) {Oct 20};
\draw[-{Stealth[length=1.8mm]},thick,cC] (m0.south) -- (j00.north);
\node[draw=cG!80,fill=white,rounded corners=2pt,minimum width=1.55cm,minimum height=1.05cm] (j10) at (1,0.75) {};
\fill[cR!75,rounded corners=1pt] (0.62,0.9600000000000001) rectangle (1.38,1.1800000000000002);
\node[font=\tiny\bfseries,white] at (1,1.07) {R};
\fill[cC!75,rounded corners=1pt] (0.62,0.64) rectangle (1.38,0.86);
\node[font=\tiny\bfseries,white] at (1,0.75) {C};
\fill[cN!75,rounded corners=1pt] (0.62,0.32000000000000006) rectangle (1.38,0.54);
\node[font=\tiny\bfseries,white] at (1,0.43000000000000005) {N};
\node[font=\tiny,cG,anchor=east] at (0.5800000000000001,0.75) {1};
\node[draw=cG!80,fill=white,rounded corners=2pt,minimum width=1.55cm,minimum height=1.05cm] (j11) at (1,-0.5) {};
\fill[cR!75,rounded corners=1pt] (0.62,-0.29) rectangle (1.38,-0.06999999999999999);
\node[font=\tiny\bfseries,white] at (1,-0.18) {R};
\fill[cC!75,rounded corners=1pt] (0.62,-0.61) rectangle (1.38,-0.39);
\node[font=\tiny\bfseries,white] at (1,-0.5) {C};
\fill[cN!75,rounded corners=1pt] (0.62,-0.93) rectangle (1.38,-0.7100000000000001);
\node[font=\tiny\bfseries,white] at (1,-0.8200000000000001) {N};
\node[font=\tiny,cG,anchor=east] at (0.5800000000000001,-0.5) {2};
\node[font=\scriptsize\bfseries] at (1,-1.55) {session 2};
\node[font=\tiny,cG] at (1,-1.85) {Oct 21};
\draw[-{Stealth[length=1.8mm]},thick,cC] (m1.south) -- (j10.north);
\node[draw=cG!80,fill=white,rounded corners=2pt,minimum width=1.55cm,minimum height=1.05cm] (j20) at (2,0.75) {};
\fill[cR!75,rounded corners=1pt] (1.62,0.9600000000000001) rectangle (2.38,1.1800000000000002);
\node[font=\tiny\bfseries,white] at (2,1.07) {R};
\fill[cC!75,rounded corners=1pt] (1.62,0.64) rectangle (2.38,0.86);
\node[font=\tiny\bfseries,white] at (2,0.75) {C};
\fill[cN!75,rounded corners=1pt] (1.62,0.32000000000000006) rectangle (2.38,0.54);
\node[font=\tiny\bfseries,white] at (2,0.43000000000000005) {N};
\node[font=\tiny,cG,anchor=east] at (1.58,0.75) {1};
\node[draw=cG!80,fill=white,rounded corners=2pt,minimum width=1.55cm,minimum height=1.05cm] (j21) at (2,-0.5) {};
\fill[cR!75,rounded corners=1pt] (1.62,-0.29) rectangle (2.38,-0.06999999999999999);
\node[font=\tiny\bfseries,white] at (2,-0.18) {R};
\fill[cC!75,rounded corners=1pt] (1.62,-0.61) rectangle (2.38,-0.39);
\node[font=\tiny\bfseries,white] at (2,-0.5) {C};
\fill[cN!75,rounded corners=1pt] (1.62,-0.93) rectangle (2.38,-0.7100000000000001);
\node[font=\tiny\bfseries,white] at (2,-0.8200000000000001) {N};
\node[font=\tiny,cG,anchor=east] at (1.58,-0.5) {2};
\node[font=\scriptsize\bfseries] at (2,-1.55) {session 3};
\node[font=\tiny,cG] at (2,-1.85) {Oct 21};
\draw[-{Stealth[length=1.8mm]},thick,cC] (m2.south) -- (j20.north);
\node[draw=cG!80,fill=white,rounded corners=2pt,minimum width=1.55cm,minimum height=1.05cm] (j30) at (3,0.75) {};
\fill[cR!75,rounded corners=1pt] (2.62,0.9600000000000001) rectangle (3.38,1.1800000000000002);
\node[font=\tiny\bfseries,white] at (3,1.07) {R};
\fill[cC!75,rounded corners=1pt] (2.62,0.64) rectangle (3.38,0.86);
\node[font=\tiny\bfseries,white] at (3,0.75) {C};
\fill[cN!75,rounded corners=1pt] (2.62,0.32000000000000006) rectangle (3.38,0.54);
\node[font=\tiny\bfseries,white] at (3,0.43000000000000005) {N};
\node[font=\tiny,cG,anchor=east] at (2.58,0.75) {1};
\node[draw=cG!80,fill=white,rounded corners=2pt,minimum width=1.55cm,minimum height=1.05cm] (j31) at (3,-0.5) {};
\fill[cR!75,rounded corners=1pt] (2.62,-0.29) rectangle (3.38,-0.06999999999999999);
\node[font=\tiny\bfseries,white] at (3,-0.18) {R};
\fill[cC!75,rounded corners=1pt] (2.62,-0.61) rectangle (3.38,-0.39);
\node[font=\tiny\bfseries,white] at (3,-0.5) {C};
\fill[cN!75,rounded corners=1pt] (2.62,-0.93) rectangle (3.38,-0.7100000000000001);
\node[font=\tiny\bfseries,white] at (3,-0.8200000000000001) {N};
\node[font=\tiny,cG,anchor=east] at (2.58,-0.5) {2};
\node[font=\scriptsize\bfseries] at (3,-1.55) {session 4};
\node[font=\tiny,cG] at (3,-1.85) {Oct 22};
\draw[-{Stealth[length=1.8mm]},thick,cC] (m3.south) -- (j30.north);
\node[draw=cG!80,fill=white,rounded corners=2pt,minimum width=1.55cm,minimum height=1.05cm] (j40) at (4,0.75) {};
\fill[cR!75,rounded corners=1pt] (3.62,0.9600000000000001) rectangle (4.38,1.1800000000000002);
\node[font=\tiny\bfseries,white] at (4,1.07) {R};
\fill[cC!75,rounded corners=1pt] (3.62,0.64) rectangle (4.38,0.86);
\node[font=\tiny\bfseries,white] at (4,0.75) {C};
\fill[cN!75,rounded corners=1pt] (3.62,0.32000000000000006) rectangle (4.38,0.54);
\node[font=\tiny\bfseries,white] at (4,0.43000000000000005) {N};
\node[font=\tiny,cG,anchor=east] at (3.58,0.75) {1};
\node[draw=cG!80,fill=white,rounded corners=2pt,minimum width=1.55cm,minimum height=1.05cm] (j41) at (4,-0.5) {};
\fill[cR!75,rounded corners=1pt] (3.62,-0.29) rectangle (4.38,-0.06999999999999999);
\node[font=\tiny\bfseries,white] at (4,-0.18) {R};
\fill[cC!75,rounded corners=1pt] (3.62,-0.61) rectangle (4.38,-0.39);
\node[font=\tiny\bfseries,white] at (4,-0.5) {C};
\fill[cN!75,rounded corners=1pt] (3.62,-0.93) rectangle (4.38,-0.7100000000000001);
\node[font=\tiny\bfseries,white] at (4,-0.8200000000000001) {N};
\node[font=\tiny,cG,anchor=east] at (3.58,-0.5) {2};
\node[font=\scriptsize\bfseries] at (4,-1.55) {session 5};
\node[font=\tiny,cG] at (4,-1.85) {Oct 23};
\draw[-{Stealth[length=1.8mm]},thick,cC] (m4.south) -- (j40.north);
\node[draw=cG!80,fill=white,rounded corners=2pt,minimum width=1.55cm,minimum height=1.05cm] (j50) at (5,0.75) {};
\fill[cR!75,rounded corners=1pt] (4.62,0.9600000000000001) rectangle (5.38,1.1800000000000002);
\node[font=\tiny\bfseries,white] at (5,1.07) {R};
\fill[cC!75,rounded corners=1pt] (4.62,0.64) rectangle (5.38,0.86);
\node[font=\tiny\bfseries,white] at (5,0.75) {C};
\fill[cN!75,rounded corners=1pt] (4.62,0.32000000000000006) rectangle (5.38,0.54);
\node[font=\tiny\bfseries,white] at (5,0.43000000000000005) {N};
\node[font=\tiny,cG,anchor=east] at (4.58,0.75) {1};
\node[draw=cG!80,fill=white,rounded corners=2pt,minimum width=1.55cm,minimum height=1.05cm] (j51) at (5,-0.5) {};
\fill[cR!75,rounded corners=1pt] (4.62,-0.29) rectangle (5.38,-0.06999999999999999);
\node[font=\tiny\bfseries,white] at (5,-0.18) {R};
\fill[cC!75,rounded corners=1pt] (4.62,-0.61) rectangle (5.38,-0.39);
\node[font=\tiny\bfseries,white] at (5,-0.5) {C};
\fill[cN!75,rounded corners=1pt] (4.62,-0.93) rectangle (5.38,-0.7100000000000001);
\node[font=\tiny\bfseries,white] at (5,-0.8200000000000001) {N};
\node[font=\tiny,cG,anchor=east] at (4.58,-0.5) {2};
\node[font=\scriptsize\bfseries] at (5,-1.55) {session 6};
\node[font=\tiny,cG] at (5,-1.85) {Oct 24};
\draw[-{Stealth[length=1.8mm]},thick,cC] (m5.south) -- (j50.north);
\node[draw=cG!80,fill=white,rounded corners=2pt,minimum width=1.55cm,minimum height=1.05cm] (j60) at (6,0.75) {};
\fill[cR!75,rounded corners=1pt] (5.62,0.9600000000000001) rectangle (6.38,1.1800000000000002);
\node[font=\tiny\bfseries,white] at (6,1.07) {R};
\fill[cC!75,rounded corners=1pt] (5.62,0.64) rectangle (6.38,0.86);
\node[font=\tiny\bfseries,white] at (6,0.75) {C};
\fill[cN!75,rounded corners=1pt] (5.62,0.32000000000000006) rectangle (6.38,0.54);
\node[font=\tiny\bfseries,white] at (6,0.43000000000000005) {N};
\node[font=\tiny,cG,anchor=east] at (5.58,0.75) {1};
\node[draw=cG!80,fill=white,rounded corners=2pt,minimum width=1.55cm,minimum height=1.05cm] (j61) at (6,-0.5) {};
\fill[cR!75,rounded corners=1pt] (5.62,-0.29) rectangle (6.38,-0.06999999999999999);
\node[font=\tiny\bfseries,white] at (6,-0.18) {R};
\fill[cC!75,rounded corners=1pt] (5.62,-0.61) rectangle (6.38,-0.39);
\node[font=\tiny\bfseries,white] at (6,-0.5) {C};
\fill[cN!75,rounded corners=1pt] (5.62,-0.93) rectangle (6.38,-0.7100000000000001);
\node[font=\tiny\bfseries,white] at (6,-0.8200000000000001) {N};
\node[font=\tiny,cG,anchor=east] at (5.58,-0.5) {2};
\node[font=\scriptsize\bfseries] at (6,-1.55) {session 7};
\node[font=\tiny,cG] at (6,-1.85) {Oct 25};
\draw[-{Stealth[length=1.8mm]},thick,cC] (m6.south) -- (j60.north);
\node[draw=cG!80,fill=white,rounded corners=2pt,minimum width=1.55cm,minimum height=1.05cm] (j70) at (7,0.75) {};
\fill[cR!75,rounded corners=1pt] (6.62,0.9600000000000001) rectangle (7.38,1.1800000000000002);
\node[font=\tiny\bfseries,white] at (7,1.07) {R};
\fill[cC!75,rounded corners=1pt] (6.62,0.64) rectangle (7.38,0.86);
\node[font=\tiny\bfseries,white] at (7,0.75) {C};
\fill[cN!75,rounded corners=1pt] (6.62,0.32000000000000006) rectangle (7.38,0.54);
\node[font=\tiny\bfseries,white] at (7,0.43000000000000005) {N};
\node[font=\tiny,cG,anchor=east] at (6.58,0.75) {1};
\node[draw=cG!80,fill=white,rounded corners=2pt,minimum width=1.55cm,minimum height=1.05cm] (j71) at (7,-0.5) {};
\fill[cR!75,rounded corners=1pt] (6.62,-0.29) rectangle (7.38,-0.06999999999999999);
\node[font=\tiny\bfseries,white] at (7,-0.18) {R};
\fill[cC!75,rounded corners=1pt] (6.62,-0.61) rectangle (7.38,-0.39);
\node[font=\tiny\bfseries,white] at (7,-0.5) {C};
\fill[cN!75,rounded corners=1pt] (6.62,-0.93) rectangle (7.38,-0.7100000000000001);
\node[font=\tiny\bfseries,white] at (7,-0.8200000000000001) {N};
\node[font=\tiny,cG,anchor=east] at (6.58,-0.5) {2};
\node[font=\scriptsize\bfseries] at (7,-1.55) {session 8};
\node[font=\tiny,cG] at (7,-1.85) {Oct 26};
\draw[-{Stealth[length=1.8mm]},thick,cC] (m7.south) -- (j70.north);
\draw[-{Stealth[length=1.5mm]},cR,dashed,line width=0.5pt,opacity=0.85] (m0.south east) to[bend left=8] (5.38,1.07);
\draw[-{Stealth[length=1.5mm]},cR,dashed,line width=0.5pt,opacity=0.85] (m2.south east) to[bend left=8] (0.38,1.07);
\draw[-{Stealth[length=1.5mm]},cR,dashed,line width=0.5pt,opacity=0.85] (m5.south east) to[bend left=8] (7.38,1.07);
\draw[-{Stealth[length=1.5mm]},cR,dashed,line width=0.5pt,opacity=0.85] (m7.south east) to[bend left=8] (3.38,1.07);
\draw[-{Stealth[length=1.5mm]},cR,dashed,line width=0.5pt,opacity=0.85] (m3.south east) to[bend left=8] (6.38,1.07);
\node[font=\scriptsize,anchor=north] at (3.5,-2.15) {\textcolor{cC}{\textbf{C}}: market day $d$ in session $d$, split in time order \quad \textcolor{cN}{\textbf{N}}: null circuit in the same slots as C \quad \textcolor{cR}{\textbf{R}}: the same events, in random jobs};
\end{tikzpicture}}
\caption{The study design. Top: the eight market days. Bottom: the eight hardware sessions, each with two jobs. Every
job carries rows from all three arms. Arm C (blue) runs market day $d$ in session $d$, split in time order. Arm N
(green) runs the null circuit in the same slots. Arm R (red) takes the same events, scattered over all sixteen jobs at
random.}
\label{fig:design}
\end{figure}

Three arms share the same 1{,}000 events, at 1{,}024 shots per measurement basis (Fig.~\ref{fig:design}).
\begin{description}
\item[R, random.] The data circuit, with each event placed in a job at random. This is the honest test of a hardware
  effect.
\item[C, natural pipeline.] The data circuit, with market day $d$ run on calibration day $d$ and split into jobs in
  time order. This is how a production pipeline would process the data.
\item[N, null.] The null circuit, in the same slots as C.
\end{description}
The runs span eight sessions, one per calibration, with two jobs per session. Every job carries rows from all three
arms, so the arms see the same hardware and differ only in which event sits in which job. A short dry run before
session 1 measures the cost per shot.

\subsection{Hardware settings}

All runs use \texttt{ibm\_fez}, a 156-qubit IBM Heron r2 processor, through the Qiskit Runtime estimator
\cite{javadi2024}. As in HSBC's study, we use Pauli twirling on gates and readout, TREX readout mitigation
\cite{vandenberg2022}, and no zero-noise extrapolation. Each job is one transpiled circuit with all of its rows as a
parameter array, measured in three bases. The sessions ran between October 19 and 26, 2026, and used 49 minutes of
processor time in total.

\subsection{Analysis and pre-registration}
\label{sec:analysis}

We use HSBC's rolling backtest. Events are taken in time order in test blocks of 40. Each block is predicted by a model
trained on the 360 most recent earlier events, ending at least the blinding window before the block. Logistic
regression and gradient boosting are tuned by cross-validation, and the better one predicts the block. We report pooled
out-of-sample AUC with blinding windows of 0, 1 and 2 days, and bootstrap paired differences over test events.

Three primary endpoints were fixed in advance, all at zero blinding:
\begin{enumerate}\setlength{\itemsep}{0pt}
\item the hardware gain, $\mathrm{AUC}(R)-\mathrm{AUC}(\text{classical})$;
\item the natural-pipeline leak, $\mathrm{AUC}(C)-\mathrm{AUC}(R)$;
\item the leak with no data, $\mathrm{AUC}(N)-0.5$.
\end{enumerate}
Each is declared positive if its 95\% interval lies above $+0.05$.

Before any run, an emulator gave predictions for each arm (Table~\ref{tab:pred}). It starts from the exact noiseless
features and adds the attenuation expected from the device's gate errors, and shot noise. It then adds per-job and
per-day offsets of unknown size $\sigma$, in units of the shot-noise standard deviation. The schedule, predictions, rules and code were
frozen with a SHA-256 fingerprint on October 19, 2026, before any hardware run. That fingerprint was not posted externally at
the time; the frozen files are deposited with this paper so that anyone can recompute it.

\begin{table}[h]
\centering\small
\begin{tabular}{lcccc}
\toprule
Arm & $\sigma=0$ & $\sigma=0.5$ & $\sigma=1$ & $\sigma=2$ \\
\midrule
R, random & 0.562 & 0.557 & 0.570 & 0.571 \\
C, natural pipeline & 0.549 & 0.563 & 0.568 & 0.563 \\
N, null circuit & 0.491 & 0.521 & 0.497 & 0.499 \\
N, null circuit, 1-day blinding & 0.477 & 0.514 & 0.485 & 0.497 \\
N replayed, sorted by outcome & 0.509 & 0.724 & 0.811 & 0.792 \\
\bottomrule
\end{tabular}
\caption{Pre-registered emulator predictions of pooled AUC at zero blinding, for job and day offsets of size $\sigma$
(shot-noise units). The baselines on the same events are classical 0.583 and noiseless 0.593.}
\label{tab:pred}
\end{table}

Two deviations from the plan occurred, both before any result was seen. The dry run measured a cost per shot slightly
above the pilot value, and the pre-registered rule then halved the null arm. Because more processor time was
available, we kept the null arm in full. And sessions 2 and 3 ran on the same two-qubit gate calibration; only the
readout was recalibrated between them. Appendix~\ref{app:deviations} gives details and shows that neither changes the
results.

%=====================================================================================================================
\section{Results}
\label{sec:results}

\subsection{The device reproduces the ideal features, attenuated}

\begin{figure}[t]
\centering
\begin{tikzpicture}\begin{axis}[width=0.62\linewidth,height=0.5\linewidth,xmin=-0.27,xmax=0.27,ymin=-0.27,ymax=0.27,xlabel={noiseless feature (centered)},ylabel={hardware feature (centered)},tick label style={font=\scriptsize},label style={font=\small},grid=major,grid style={gray!15},legend style={font=\scriptsize,at={(0.03,0.97)},anchor=north west,draw=none,fill=white,fill opacity=0.85,text opacity=1}]
\addplot[only marks,mark=*,mark size=1.1pt,cR,fill opacity=0.35,draw opacity=0.5] coordinates {(-0.0209,-0.0506) (-0.0478,0.0418) (0.0751,-0.0705) (0.0595,-0.0116) (-0.0114,-0.1269) (-0.0011,-0.0612) (0.0743,0.0867) (0.0426,-0.0312) (-0.0862,-0.0827) (0.0399,0.1732) (0.0352,-0.0713) (0.0197,-0.0222) (-0.0288,0.1415) (-0.087,-0.1404) (-0.0755,0.1077) (0.1888,0.1145) (-0.056,-0.0633) (0.0251,0.1034) (0.1281,0.0973) (-0.0531,-0.0549) (-0.0737,-0.0033) (0.018,0.1013) (0.0063,0.0103) (-0.0582,0.188) (0.1213,0.2455) (0.0227,0.148) (-0.0244,-0.1847) (-0.0388,-0.1233) (-0.005,-0.0097) (0.2402,0.0314) (-0.1212,0.11) (-0.0506,-0.1212) (0.0674,0.0204) (-0.0428,0.0227) (0.0583,-0.0136) (0.04,-0.0547) (0.0094,0.0442) (0.0917,0.0374) (0.0266,-0.0204) (-0.0256,-0.2314) (-0.1265,0.1036) (0.1049,-0.0555) (-0.0581,-0.0571) (-0.1369,-0.108) (-0.028,0.176) (-0.0161,0.0396) (0.1073,0.1755) (-0.0863,0.0103) (-0.1396,-0.1775) (0.0704,-0.0656) (-0.0259,-0.1162) (-0.0308,-0.1508) (0.1141,-0.0508) (0.2316,0.1531) (-0.1272,-0.0065) (0.0914,0.0653) (-0.034,0.237) (0.0482,0.076) (-0.032,-0.1766) (0.0753,-0.0022) (-0.1066,-0.1459) (-0.0509,-0.112) (-0.0218,0.0035) (-0.0014,0.0041) (0.0832,0.061) (0.0232,-0.0065) (-0.0824,0.1565) (-0.0727,-0.0655) (-0.0268,0.0227) (-0.0808,-0.1966) (0.1067,0.0898) (-0.0653,0.0185) (0.0507,-0.1015) (0.051,0.0214) (0.0443,0.0956) (0.1111,-0.0631) (-0.0567,-0.0388) (-0.0547,-0.0238) (0.0218,0.0876) (0.0099,-0.0929) (0.0006,-0.0607) (0.0341,0.0359) (0.1106,0.0776) (-0.1292,0.018) (-0.0726,-0.1272) (0.0446,0.0452) (0.0925,0.079) (0.0422,0.0031) (0.0893,-0.0953) (-0.0249,-0.0399) (0.1407,-0.0125) (0.0083,0.0392) (-0.0891,0.0394) (-0.0612,0.0092) (0.1404,-0.0733) (-0.0238,0.0045) (0.0143,0.0704) (0.0186,0.0334) (0.1031,0.0704) (0.049,0.0151) (-0.1323,-0.0209) (-0.1281,-0.1432) (0.0096,0.0013) (0.0074,0.0616) (-0.015,-0.0278) (0.1335,-0.0576) (-0.1152,-0.0853) (-0.0657,-0.0891) (0.0608,0.1037) (-0.0146,0.0297) (0.1763,0.1174) (0.0026,-0.0953) (0.0934,-0.1133) (0.0224,-0.0671) (-0.1272,-0.0079) (-0.1569,-0.1113) (0.0595,0.0552) (-0.0248,-0.0733) (-0.1191,-0.0234) (0.0059,-0.0337) (0.0372,0.0773) (-0.0127,-0.0728) (-0.0422,0.0339) (0.0053,0.0079) (0.0233,0.0391) (0.0426,-0.0286) (-0.0719,-0.0377) (0.1062,0.0239) (0.0076,0.0114) (-0.0274,-0.005) (0.0298,0.0378) (0.0121,0.0099) (0.0988,0.0776) (-0.1171,-0.0278) (-0.0505,0.0059) (-0.1574,-0.0578) (0.0448,0.071) (-0.0563,-0.0067) (0.0939,-0.0417) (0.0973,0.0752) (-0.015,-0.087) (-0.0027,0.0231) (-0.0404,-0.0381) (-0.1345,-0.0523) (-0.048,-0.0081) (-0.0399,0.0349) (-0.1022,-0.0276) (-0.0799,-0.0712) (-0.0202,0.0007) (0.1246,-0.0259) (-0.0487,-0.0362) (-0.0887,0.025) (-0.0592,0.0767) (-0.0083,-0.0018) (-0.2035,-0.1196) (-0.0674,-0.0263) (-0.0019,0.031) (-0.1422,-0.0454) (0.1009,0.0254) (-0.0094,-0.0503) (0.0109,0.0112) (0.0413,-0.0414) (0.0388,0.002) (0.0719,-0.046) (0.0359,0.0836) (-0.0023,0.012) (0.1285,0.028) (0.082,0.0767) (0.0916,0.039) (-0.0276,-0.0873) (-0.0241,0.0333) (0.0671,0.057) (0.09,0.0745) (-0.0099,0.0171) (0.0174,-0.0514) (0.1163,0.0799) (0.1036,0.0398) (-0.0373,-0.0158) (0.0149,-0.0066) (0.0177,-0.0009) (-0.0322,-0.0499) (-0.101,0.0669) (-0.0772,-0.0263) (-0.1787,-0.0954) (0.0791,0.0379) (-0.0305,-0.0039) (0.0081,0.01) (-0.02,-0.0037) (0.0148,-0.012) (0.0241,-0.0464) (0.0675,0.0627) (0.0626,0.006) (-0.0983,-0.0385) (0.1325,-0.0337) (-0.1946,-0.0607) (-0.0032,-0.0183) (0.0438,0.0389) (0.0431,-0.0219) (0.0598,0.0355) (0.0607,0.0153) (0.0177,-0.0077) (-0.0047,-0.0032) (-0.0496,0.0314) (0.1426,0.0215) (0.0978,0.063) (0.0063,-0.0346) (-0.0836,-0.0243) (-0.0381,0.0412) (-0.1317,-0.048) (0.0141,-0.0306) (-0.1821,-0.0977) (-0.0394,-0.044) (0.0084,0.0193) (0.0865,0.023) (-0.2158,-0.1089) (-0.083,-0.04) (-0.0888,-0.0538) (0.0298,0.0193) (0.1121,0.023) (-0.1256,-0.0936) (-0.0404,-0.1208) (0.0459,0.1134) (0.0563,0.0133) (-0.0946,-0.1072) (0.0177,-0.0184) (-0.132,-0.0776) (0.0333,0.0431) (-0.05,-0.042) (0.0128,0.0347) (0.0315,0.0015) (-0.0257,0.0485) (0.0145,0.0113) (-0.0892,-0.0853) (-0.042,-0.0083) (-0.0031,0.0269) (-0.0899,-0.05) (0.0246,0.0151) (0.0887,0.0289) (-0.0345,0.0391) (0.0837,0.1277) (-0.1539,-0.1604) (0.0969,0.0411) (-0.0662,-0.0003) (-0.0704,-0.0893) (0.0182,-0.0164) (-0.0907,0.0135) (-0.0679,-0.044) (-0.0658,0.076) (-0.14,-0.1385) (0.1001,0.075) (-0.0807,-0.0932) (-0.0443,-0.0578) (-0.056,-0.0302) (-0.0639,-0.0164) (-0.1235,-0.117) (-0.1885,-0.0757) (-0.0294,-0.0222) (-0.0105,0.0249) (-0.1177,-0.0617) (-0.075,0.0073) (-0.0791,-0.0479) (0.0299,-0.0912) (-0.0412,0.0485) (-0.0361,0.0705) (-0.074,-0.0103) (0.0008,-0.0894) (-0.0627,-0.0715) (-0.0835,-0.0459) (0.0575,0.0527) (0.1063,0.1254) (-0.079,-0.048) (0.0777,0.0726) (0.0084,0.0135) (0.0769,-0.0007) (0.0946,-0.0024) (0.0922,0.1016) (-0.1241,-0.0302) (0.0166,0.037) (-0.0368,0.0465) (-0.0723,-0.0538) (0.0449,0.0224) (0.0323,-0.0001) (0.0594,0.0361) (0.0885,0.0616) (0.0419,0.03) (-0.0418,-0.0818) (-0.0171,-0.037) (-0.0652,-0.032) (-0.0222,-0.0747) (-0.0691,-0.0083) (0.0612,0.0489) (-0.1482,-0.1498) (0.0222,0.0228) (-0.1393,-0.061) (0.0344,-0.0079) (0.095,0.034) (0.0069,-0.023) (0.053,0.0165) (0.0268,-0.0454) (0.0607,-0.0107) (0.1109,0.1339) (-0.112,-0.0995) (-0.113,-0.0766) (-0.0092,0.0206) (-0.0519,-0.0791) (0.0191,0.0293) (0.0873,0.0595) (0.0318,0.0313) (0.0023,0.0065) (0.0248,-0.0211) (-0.0803,-0.086) (0.013,-0.0298) (-0.0952,-0.0672) (0.009,-0.049) (-0.0734,-0.0889) (-0.0677,-0.0596) (0.0041,0.0744) (0.0854,0.0343) (0.0464,0.0442) (0.041,0.0434) (0.0863,0.0687) (-0.053,-0.0379) (-0.0618,-0.025) (0.0387,0.0008) (-0.0779,0.0045) (0.0719,0.0215) (0.0065,-0.032) (0.0154,0.0124) (0.0559,0.106) (-0.0865,-0.1334) (0.0232,-0.0253) (0.0735,0.0382) (0.0029,0.0272) (0.1158,0.085) (0.0504,0.0616) (-0.152,-0.1123) (-0.1321,-0.144) (0.0169,-0.0553) (-0.0009,0.0479) (-0.0496,-0.0171) (0.01,0.0399) (-0.1725,-0.0974) (-0.0798,-0.0612) (0.0503,0.1413) (-0.0893,-0.0213) (0.1381,0.0892) (-0.0513,-0.0405) (0.1051,0.0378) (-0.0716,-0.0753) (-0.0939,-0.1101) (-0.0408,-0.0014) (-0.0809,-0.0939) (0.0092,0.0028) (0.0059,0.0121) (0.0011,0.0799) (-0.0303,-0.0588) (-0.0345,0.0148) (-0.0752,-0.0549) (-0.0588,-0.0132) (0.0046,0.0026) (-0.0617,-0.0375) (-0.0322,-0.0394) (-0.0768,-0.0508) (-0.0985,-0.0999) (-0.04,-0.0436) (-0.0696,0.0301) (0.0177,0.0782) (0.0966,0.0588) (0.0148,0.0516) (0.1285,0.0599) (-0.0556,0.0437) (0.0485,0.0101) (-0.04,-0.0885) (-0.0185,0.0179) (0.0115,0.0484) (-0.0023,0.0424) (0.0784,0.0806) (-0.0117,-0.0246) (0.0802,0.0737) (-0.0688,-0.0207) (-0.0266,0.0168) (-0.0053,0.0462) (0.1781,0.0934) (-0.0262,-0.0039) (0.0107,0.0042) (-0.0173,-0.0294) (-0.0869,-0.071) (-0.0385,-0.0649) (0.2018,0.0948) (0.1536,0.0663) (0.0232,-0.0215) (-0.0151,0.0208) (0.0916,0.0918) (-0.1112,-0.084) (0.0186,0.0501) (-0.0508,-0.0426) (0.0489,0.0655) (-0.0582,-0.0248) (0.0082,0.0397) (-0.034,0.0101) (0.1191,0.0873) (0.0631,0.038) (0.0552,0.0219) (-0.0079,-0.0708) (0.102,0.0832) (-0.0411,0.0161) (-0.0553,-0.0787) (0.0284,-0.0116) (-0.0345,0.0126) (-0.0033,0.0167) (-0.1127,-0.009) (-0.0079,-0.0294) (-0.0625,-0.0268) (-0.035,-0.0217) (-0.0653,-0.059) (-0.0804,-0.0847) (-0.0873,-0.095) (0.0382,0.0503) (0.0377,0.0575) (-0.0534,-0.0945)};
\addlegendentry{one event, one feature}
\addplot[domain=-0.27:0.27,samples=2,cR,line width=1.4pt] {0.616*x};
\addlegendentry{median slope 0.62}
\addplot[domain=-0.27:0.27,samples=2,cG,dashed,line width=1pt] {x};
\addlegendentry{perfect agreement}
\end{axis}\end{tikzpicture}
\caption{Hardware against noiseless features (arm R). Each point is one event and one of the six features with the
largest ideal variation, both centered. The red line is the median slope over all features; the dashed line is
perfect agreement.}
\label{fig:faithful}
\end{figure}
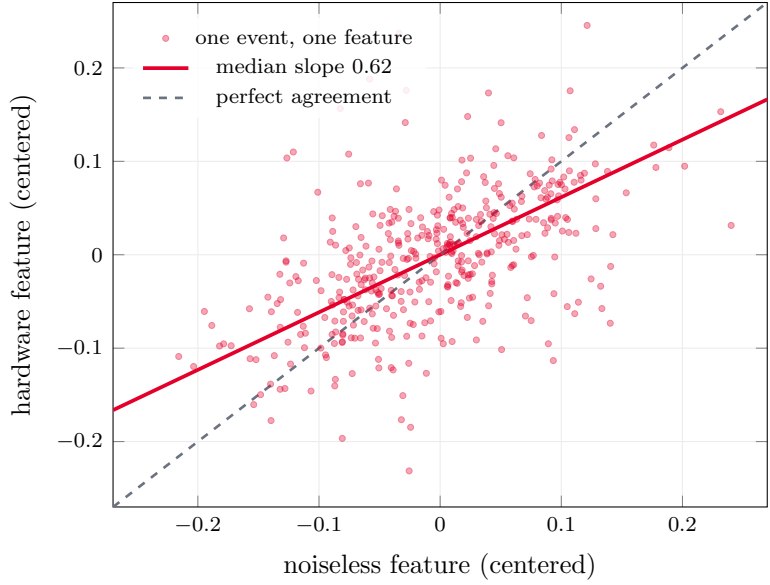

Figure~\ref{fig:faithful} compares hardware and noiseless features for the same events. The device keeps about
62\% of the ideal variation across events and adds noise that is almost exactly shot noise. The correlation
between hardware and ideal features (median 0.39 over features) sits close to the ceiling that shot noise alone would
allow (0.42). In other words, at 129 qubits and 3{,}072 CZ gates, the hardware features are an attenuated, noisy copy
of the noiseless ones. Nothing in them looks like extra structure.

\subsection{No gain under controlled assignment}

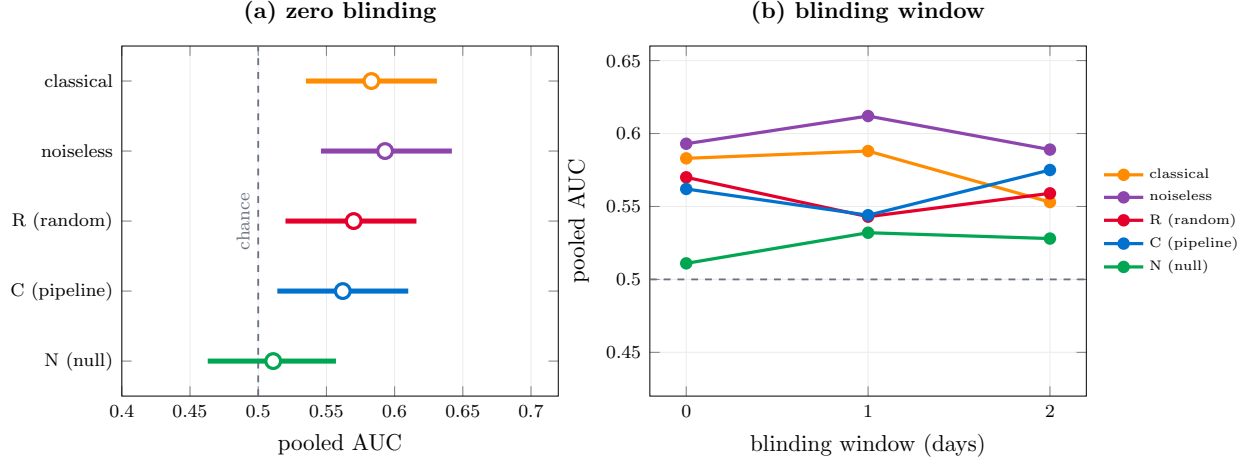
\begin{figure}[t]
\centering
\resizebox{\linewidth}{!}{\begin{tikzpicture}\begin{axis}[name=a,width=0.5\linewidth,height=0.42\linewidth,xmin=0.4,xmax=0.72,ymin=0.5,ymax=5.5,ytick={1,...,5},yticklabels={N (null),C (pipeline),R (random),noiseless,classical},xlabel={pooled AUC},tick label style={font=\scriptsize},label style={font=\small},xmajorgrids,grid style={gray!15},title={(a) zero blinding},title style={font=\small\bfseries}]
\draw[cG,dashed,thick] (axis cs:0.5,0.5) -- (axis cs:0.5,5.5);
\node[font=\scriptsize,cG,rotate=90,anchor=south] at (axis cs:0.5,3) {chance};
\draw[cN,line width=2.2pt] (axis cs:0.463,1) -- (axis cs:0.557,1);
\addplot[only marks,mark=*,mark size=3.2pt,cN,mark options={fill=white,line width=1.4pt}] coordinates {(0.511,1)};
\draw[cC,line width=2.2pt] (axis cs:0.514,2) -- (axis cs:0.61,2);
\addplot[only marks,mark=*,mark size=3.2pt,cC,mark options={fill=white,line width=1.4pt}] coordinates {(0.562,2)};
\draw[cR,line width=2.2pt] (axis cs:0.52,3) -- (axis cs:0.616,3);
\addplot[only marks,mark=*,mark size=3.2pt,cR,mark options={fill=white,line width=1.4pt}] coordinates {(0.57,3)};
\draw[cQ,line width=2.2pt] (axis cs:0.546,4) -- (axis cs:0.642,4);
\addplot[only marks,mark=*,mark size=3.2pt,cQ,mark options={fill=white,line width=1.4pt}] coordinates {(0.593,4)};
\draw[cK,line width=2.2pt] (axis cs:0.535,5) -- (axis cs:0.631,5);
\addplot[only marks,mark=*,mark size=3.2pt,cK,mark options={fill=white,line width=1.4pt}] coordinates {(0.583,5)};
\end{axis}
\begin{axis}[at={(a.east)},anchor=west,xshift=1.4cm,width=0.5\linewidth,height=0.42\linewidth,xmin=-0.2,xmax=2.2,xtick={0,1,2},ymin=0.42,ymax=0.66,xlabel={blinding window (days)},ylabel={pooled AUC},tick label style={font=\scriptsize},label style={font=\small},grid=major,grid style={gray!15},title={(b) blinding window},title style={font=\small\bfseries},legend style={font=\tiny,at={(1.02,0.5)},anchor=west,draw=none},legend cell align=left]
\draw[cG,dashed,thick] (axis cs:-0.2,0.5) -- (axis cs:2.2,0.5);
\addplot[cK,line width=1.4pt,mark=*,mark size=2pt] coordinates {(0,0.583) (1,0.588) (2,0.553)};
\addlegendentry{classical}
\addplot[cQ,line width=1.4pt,mark=*,mark size=2pt] coordinates {(0,0.593) (1,0.612) (2,0.589)};
\addlegendentry{noiseless}
\addplot[cR,line width=1.4pt,mark=*,mark size=2pt] coordinates {(0,0.570) (1,0.543) (2,0.559)};
\addlegendentry{R (random)}
\addplot[cC,line width=1.4pt,mark=*,mark size=2pt] coordinates {(0,0.562) (1,0.544) (2,0.575)};
\addlegendentry{C (pipeline)}
\addplot[cN,line width=1.4pt,mark=*,mark size=2pt] coordinates {(0,0.511) (1,0.532) (2,0.528)};
\addlegendentry{N (null)}
\end{axis}\end{tikzpicture}}
\caption{(a)~Pooled out-of-sample AUC at zero blinding, with 95\% bootstrap intervals, on the same 1{,}000 events.
(b)~AUC against the blinding window between training and test.}
\label{fig:endpoints}
\end{figure}

Figure~\ref{fig:endpoints} and Table~\ref{tab:endpoints} give the endpoints. Hardware features under random
assignment do slightly worse than classical features, and the interval excludes a gain of $+0.05$, let alone the
reported one. The natural pipeline adds nothing over random assignment. The null circuit sits at chance. The same holds
at blinding windows of one and two days. The exact noiseless features do not beat classical features either, so there
is no signal for the device to lose or to amplify.

\begin{table}[h]
\centering\small
\begin{tabular}{lccc}
\toprule
Endpoint (zero blinding) & Estimate & 95\% interval & Declared \\
\midrule
1. Hardware gain, $\mathrm{AUC}(R)-\mathrm{AUC}(\text{classical})$ & $-0.013$ & $-0.075$ to $+0.047$ & no \\
2. Natural-pipeline leak, $\mathrm{AUC}(C)-\mathrm{AUC}(R)$ & $-0.008$ & $-0.071$ to $+0.057$ & no \\
3. Leak with no data, $\mathrm{AUC}(N)-0.5$ & $+0.011$ & $-0.037$ to $+0.057$ & no \\
\bottomrule
\end{tabular}
\caption{Pre-registered primary endpoints, each paired over 640 test events.}
\label{tab:endpoints}
\end{table}

\subsection{Every job carries a fingerprint}

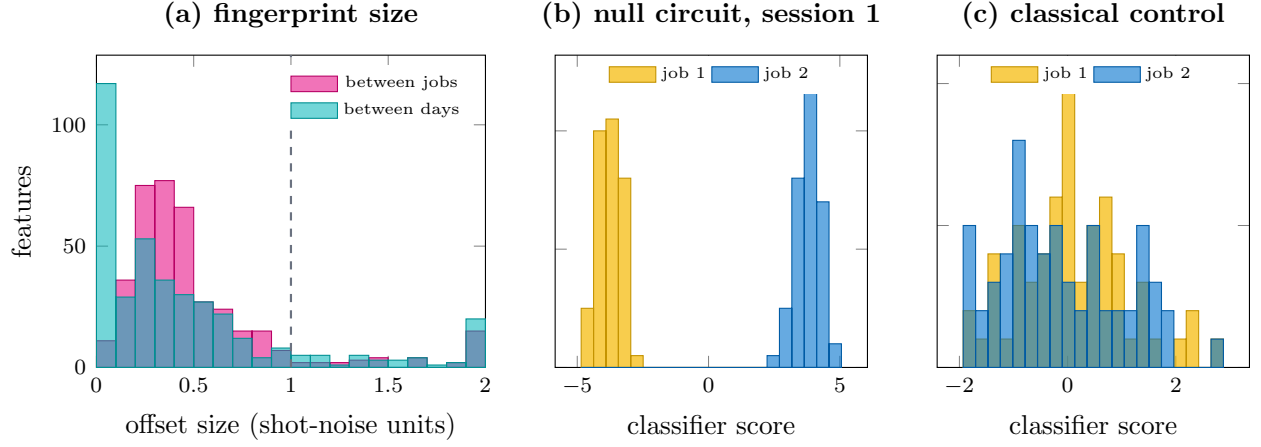
\begin{figure}[t]
\centering
\resizebox{\linewidth}{!}{\begin{tikzpicture}\begin{axis}[name=a,width=0.4\linewidth,height=0.34\linewidth,xmin=0,xmax=2,ymin=0,xlabel={offset size (shot-noise units)},ylabel={features},tick label style={font=\scriptsize},label style={font=\small},title={(a) fingerprint size},title style={font=\small\bfseries},legend style={font=\tiny,draw=none,at={(0.97,0.97)},anchor=north east}]
\addplot[ybar interval,area legend,fill=cM,fill opacity=0.55,draw=cM!80!black] coordinates {(0.000,11) (0.100,36) (0.200,75) (0.300,77) (0.400,66) (0.500,27) (0.600,24) (0.700,15) (0.800,15) (0.900,7) (1.000,2) (1.100,2) (1.200,2) (1.300,3) (1.400,4) (1.500,0) (1.600,4) (1.700,0) (1.800,2) (1.900,15) (2.000,0)};
\addlegendentry{between jobs}
\addplot[ybar interval,area legend,fill=cT,fill opacity=0.55,draw=cT!80!black] coordinates {(0.000,117) (0.100,29) (0.200,53) (0.300,36) (0.400,30) (0.500,27) (0.600,22) (0.700,12) (0.800,4) (0.900,8) (1.000,5) (1.100,5) (1.200,1) (1.300,5) (1.400,3) (1.500,3) (1.600,4) (1.700,1) (1.800,2) (1.900,20) (2.000,0)};
\addlegendentry{between days}
\draw[cG,dashed,thick] (axis cs:1,0) -- (axis cs:1,\pgfkeysvalueof{/pgfplots/ymax});
\end{axis}
\begin{axis}[name=b,at={(a.east)},anchor=west,xshift=0.9cm,width=0.34\linewidth,height=0.34\linewidth,ymin=0,xlabel={classifier score},yticklabels={},tick label style={font=\scriptsize},label style={font=\small},title={(b) null circuit, session 1},title style={font=\small\bfseries},legend style={font=\tiny,draw=none,at={(0.5,1.0)},anchor=north},legend columns=2]
\addplot[ybar interval,area legend,fill=cY,fill opacity=0.7,draw=cY!80!black] coordinates {(-4.850,5) (-4.378,20) (-3.906,21) (-3.434,16) (-2.962,1) (-2.490,0) (-2.019,0) (-1.547,0) (-1.075,0) (-0.603,0) (-0.131,0) (0.341,0) (0.813,0) (1.285,0) (1.757,0) (2.229,0) (2.700,0) (3.172,0) (3.644,0) (4.116,0) (4.588,0) (5.060,0)};
\addlegendentry{job 1}
\addplot[ybar interval,area legend,fill=cC,fill opacity=0.6,draw=cC!80!black] coordinates {(-4.850,0) (-4.378,0) (-3.906,0) (-3.434,0) (-2.962,0) (-2.490,0) (-2.019,0) (-1.547,0) (-1.075,0) (-0.603,0) (-0.131,0) (0.341,0) (0.813,0) (1.285,0) (1.757,0) (2.229,1) (2.700,5) (3.172,16) (3.644,24) (4.116,14) (4.588,2) (5.060,0)};
\addlegendentry{job 2}
\end{axis}
\begin{axis}[name=c,at={(b.east)},anchor=west,xshift=0.9cm,width=0.34\linewidth,height=0.34\linewidth,ymin=0,xlabel={classifier score},yticklabels={},tick label style={font=\scriptsize},label style={font=\small},title={(c) classical control},title style={font=\small\bfseries},legend style={font=\tiny,draw=none,at={(0.5,1.0)},anchor=north},legend columns=2]
\addplot[ybar interval,area legend,fill=cY,fill opacity=0.7,draw=cY!80!black] coordinates {(-1.930,2) (-1.701,1) (-1.472,4) (-1.243,1) (-1.014,5) (-0.785,3) (-0.556,4) (-0.327,6) (-0.098,10) (0.131,3) (0.360,5) (0.590,6) (0.819,4) (1.048,1) (1.277,3) (1.506,1) (1.735,0) (1.964,1) (2.193,2) (2.422,0) (2.651,1) (2.880,0)};
\addlegendentry{job 1}
\addplot[ybar interval,area legend,fill=cC,fill opacity=0.6,draw=cC!80!black] coordinates {(-1.930,5) (-1.701,2) (-1.472,3) (-1.243,4) (-1.014,8) (-0.785,5) (-0.556,4) (-0.327,5) (-0.098,3) (0.131,2) (0.360,5) (0.590,2) (0.819,2) (1.048,2) (1.277,5) (1.506,3) (1.735,2) (1.964,0) (2.193,0) (2.422,0) (2.651,1) (2.880,0)};
\addlegendentry{job 2}
\end{axis}
\end{tikzpicture}}
\caption{(a)~Size of the job and day offsets of each of the 387 null-arm features, in shot-noise units. Offsets too
small to tell from zero fall in the first bin, and values above 2 in the last. (b)~Out-of-fold
classifier scores for the two jobs of session 1, from null-circuit rows: the jobs separate completely. (c)~The same
classifier on the classical features of the same session's rows: the jobs overlap, as they should.}
\label{fig:fingerprints}
\end{figure}

The null arm measures the device's fingerprint directly. Within a job, null rows scatter almost exactly as shot
noise predicts. Between the two jobs of a session, the mean of a typical feature shifts by about 0.4 shot-noise widths, and
between days by about 0.3 (Fig.~\ref{fig:fingerprints}a). These shifts are small for any one feature. Summed over 387
features they are not. A cross-validated classifier tells the two jobs of every session apart perfectly, from null rows
and from data rows alike. It also names the session of a null row correctly every time. The classical features of the same
rows stay at chance (Fig.~\ref{fig:fingerprints}b,c). The fingerprints are also random: the difference between a
session's two jobs does not repeat from one session to the next. Each job has its own barcode.

\subsection{Batching by outcome fakes a signal, and purging removes it}

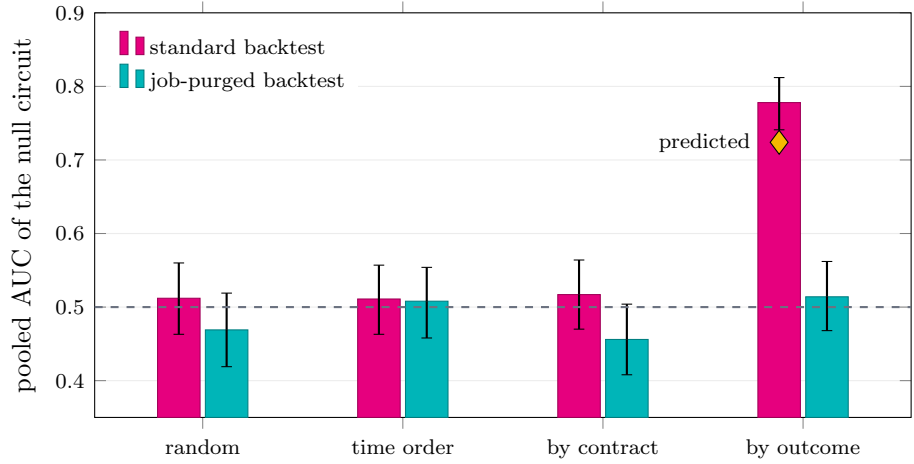
\begin{figure}[t]
\centering
\begin{tikzpicture}\begin{axis}[width=0.75\linewidth,height=0.42\linewidth,ybar=2pt,bar width=16pt,ymin=0.35,ymax=0.9,xtick=data,symbolic x coords={random,time order,by contract,by outcome},enlarge x limits=0.18,ylabel={pooled AUC of the null circuit},tick label style={font=\scriptsize},label style={font=\small},ymajorgrids,grid style={gray!15},legend style={font=\scriptsize,at={(0.02,0.97)},anchor=north west,draw=none},legend cell align=left]
\addplot[fill=cM,draw=cM!70!black,error bars/.cd,y dir=both,y explicit,error bar style={black,line width=0.8pt}] coordinates {(random,0.512) += (0,0.048) -= (0,0.049) (time order,0.511) += (0,0.046) -= (0,0.048) (by contract,0.517) += (0,0.047) -= (0,0.047) (by outcome,0.778) += (0,0.034) -= (0,0.037)};
\addplot[fill=cT,draw=cT!70!black,error bars/.cd,y dir=both,y explicit,error bar style={black,line width=0.8pt}] coordinates {(random,0.469) += (0,0.050) -= (0,0.050) (time order,0.508) += (0,0.046) -= (0,0.050) (by contract,0.456) += (0,0.048) -= (0,0.048) (by outcome,0.514) += (0,0.048) -= (0,0.046)};
\legend{standard backtest,job-purged backtest}
\draw[cG,dashed,thick] ({rel axis cs:0,0}|-{axis cs:random,0.5}) -- ({rel axis cs:1,0}|-{axis cs:random,0.5});
\addplot[only marks,mark=diamond*,mark size=4.5pt,cY,mark options={draw=black},forget plot,xshift=-9pt] coordinates {(by outcome,0.724)};
\node[font=\scriptsize,anchor=east] at ([xshift=-16pt]axis cs:by outcome,0.724) {predicted};
\end{axis}\end{tikzpicture}
\caption{Replay of the null-circuit rows under four submission orders, with the standard and the job-purged
backtest. Bars give the pooled AUC with 95\% intervals; the diamond marks the pre-registered prediction for
outcome-sorted batching.}
\label{fig:replay}
\end{figure}

The replay asks what each submission order would have produced on this device with no data at all
(Fig.~\ref{fig:replay}). Random assignment, time order and grouping by contract all leave the null circuit at chance.
Sorting each day's events by outcome before batching gives the null circuit an AUC near 0.78, close to the
pre-registered prediction. A job-purged backtest brings it back to chance. Removing each job's mean has the same
effect, and so does a one-day blinding window, because none of our jobs spans two days. The fake signal is the job
barcode, nothing more.

\subsection{Measured against predicted}

The emulator predicted every arm before the runs, for a range of fingerprint sizes. The measured job fingerprint lies
near the $\sigma=0.5$ column of Table~\ref{tab:pred}, and every arm lands within about 0.05 of its prediction
there: R at 0.570 (predicted 0.557), C at 0.562 (0.563), N at 0.511 (0.521), and the outcome-sorted replay
at 0.78 (0.724). The study behaved as designed.

%=====================================================================================================================
\section{Discussion}
\label{sec:discussion}

\paragraph{What this says about the HSBC/IBM result.} We did not reproduce a hardware gain, on a public task with the
same structure, the same circuit at a larger scale, and the same mitigation. We also found a mechanism that can produce
a large apparent gain from a circuit that carries no data. It needs two ingredients: jobs whose records share an
outcome, and a backtest in which the same job sits on both sides of the split. A rolling backtest with no blinding
window supplies the second ingredient whenever a job's records straddle a test block. We cannot say how HSBC's records were grouped into jobs,
and we do not claim that this is what happened. We do claim three things. The mechanism is sufficient. It matches the
reported decay with the blinding window. And the protocol can rule it in or out with data the original study already
had.

\paragraph{The fingerprint is the central fact.} The per-feature offsets are well below the shot noise, which is why
they are easy to dismiss. But a model reads all features at once, and hundreds of small offsets make a barcode that a
linear classifier reads perfectly. Our pipeline did not leak because R and C did not group records by outcome. Real
pipelines can: data sorted by date and then by label, or grouped by client or venue, with outcomes that differ by
group.
At higher shot counts the shot noise shrinks while the device drift should not, so the barcode should become sharper
still.

\paragraph{Limitations.} We tested one task, one circuit family and one device. The task is an analog of HSBC's, not
their data. We ran 1{,}000 events at 1{,}024 shots, against their 16{,}000 at 4{,}096. Our intervals are therefore wider. They
still exclude gains well below the one reported. Our primary models were logistic regression and gradient boosting,
not HSBC's full four-family grids. Since even exact noiseless features give no gain over classical ones, heavier
tuning has nothing to find. The pre-registration was frozen before the runs, but its fingerprint was not posted
externally until publication. Two deviations from the plan are reported in Appendix~\ref{app:deviations}.

\paragraph{A checklist for hardware feature claims.} Report how records were assigned to jobs. Run a null circuit in
the same jobs. Show that jobs cannot be identified from the features, or that the pipeline does not let job identity
correlate with the outcome. Compare random and natural assignment on the same records. Use a job-purged backtest, and
report results at more than one blinding window. Fix the analysis in advance.

%=====================================================================================================================
\section{Conclusion}

A quantum processor is a measuring instrument with a memory of when it was used. That memory is invisible in any single
feature and obvious across hundreds of them. It is enough to make a circuit with no data look like a good predictor,
if the pipeline lets it. The protocol proposed here is a small set of cheap checks that tell a real hardware effect
from this kind of leak. Applied to a prominent claim, it found no quantum gain and a clear account of how one could
appear. We hope it becomes a routine part of hardware machine-learning studies on tabular data.

%=====================================================================================================================
\section*{Acknowledgments}
We acknowledge the use of IBM Quantum services for this work. The views expressed are those of the author and do not
reflect the official policy or position of IBM or the IBM Quantum team. This work was supported by the Public Sector
Development Programme (PSDP) grant for the nanoscience center at LUMS.

\section*{Use of AI tools}
The author used Claude (Anthropic), an AI assistant, to help write and test the analysis code and to edit the
manuscript. The author designed the study, made the design decisions, ran the hardware sessions, checked the results
and takes full responsibility for the content.

\section*{Data and code availability}
Code, notebooks, the frozen pre-registration and the results are at
\url{https://github.com/muf148/hardware-qml-audit} (archived as \codeDOI). The hardware outputs for every job, the
device calibration at each session and the event and feature tables are on Zenodo (\dataDOI). The market data are
public \cite{binance}.

%=====================================================================================================================
\appendix
\section{Deviations from the plan}
\label{app:deviations}

\paragraph{The null arm was kept in full.} The dry run billed its two jobs in whole seconds. The fitted cost was
305\,$\mu$s per shot, about 7\% above the pilot value, and projected a total just above the 50-minute cap. The
pre-registered rule then halved the null arm. Additional processor time was available, so before session 1, and with no
result seen, we kept the null arm at 1{,}000 events. The decision and the rule's original output are both recorded in
the hardware log. With larger jobs the cost settled at 297\,$\mu$s per shot and 10.5\,s per job, and the study used
49.1 minutes.

\paragraph{Two sessions shared a gate calibration.} The device's calibration stamp changes several times a day, but most
updates re-measure only the readout. Our scheduler counted a new stamp as a new calibration. Sessions 2 and 3 therefore
ran nine hours apart on the same two-qubit gate calibration, with only 2\% of the chain's CZ gates re-measured between
them. Every other pair of consecutive sessions had more than 95\% of them re-measured, and the scheduler was changed to
require this from session 5 on. The null-arm means of sessions 2 and 3 differ from each other as much as those of other
consecutive sessions, and merging the two sessions changes the day-level fingerprint only slightly
(Table~\ref{tab:exploratory}).

\section{Exploratory checks}
\label{app:exploratory}

Table~\ref{tab:exploratory} lists the checks added after the runs. None is an endpoint.

\begin{table}[h]
\centering\small
\begin{tabularx}{\linewidth}{@{}>{\raggedright\arraybackslash}p{3.4cm}>{\raggedright\arraybackslash}X r@{}}
\toprule
Check & Quantity & Value \\
\midrule
Device against noiseless (arm R) & correlation with noiseless, median over features & 0.39 \\
 & same, if the device only added shot noise & 0.42 \\
 & residual scatter / shot noise & 1.05 \\
\addlinespace[2pt]
Which job of a session? & null rows: AUC (sessions with $p<0.05$) & 1.00 (8 of 8) \\
 & data rows, arm R & 1.00 (8 of 8) \\
 & classical features of the same rows (control) & 0.53 (1 of 8) \\
\addlinespace[2pt]
Which session? (chance 0.125) & null rows; data rows; control: accuracy & 1.00; 1.00; 0.12 \\
\addlinespace[2pt]
Job barcodes & correlation of job differences across sessions & $-0.02$ \\
 & separation of a session's two jobs, $d'$ & 17 to 24 \\
\addlinespace[2pt]
Outcome-sorted replay & standard backtest & 0.778 [0.741, 0.812] \\
 & job-purged backtest & 0.514 [0.468, 0.562] \\
 & job means removed & 0.503 [0.457, 0.549] \\
 & 1-day blinding & 0.487 [0.430, 0.543] \\
 & 2-day blinding & 0.440 [0.381, 0.502] \\
\addlinespace[2pt]
Paired AUC differences & noiseless $-$ classical & $+0.010$ [$-0.037$, $+0.056$] \\
 & R $-$ noiseless & $-0.023$ [$-0.073$, $+0.026$] \\
 & (classical $+$ R) $-$ classical & $+0.023$ [$-0.019$, $+0.066$] \\
\addlinespace[2pt]
Deviation 2 & null-arm mean shift, sessions 2 vs 3; other pairs & 0.0247; 0.0241 \\
 & day offset, median; sessions 2 and 3 merged & 0.29; 0.27 \\
\bottomrule
\end{tabularx}
\caption{Exploratory checks (not pre-registered). AUC differences are paired over the same test events.}
\label{tab:exploratory}
\end{table}

%=====================================================================================================================

\end{document}